\documentclass[11pt, draft, onecolumn]{IEEEtran}
\usepackage{xspace,amsmath, amssymb,epsfig,syntonly,color,enumerate,bm,bbm}
\usepackage{xcolor}
\usepackage{amsfonts}
\usepackage{graphicx}
\usepackage[lofdepth,lotdepth]{subfig}

\psfull

\newcommand{\beq} {\begin{equation}}
\newcommand{\eeq} {\end{equation}}

\newtheorem{proposition}{\hspace{-1pt}\bf Proposition}
\newtheorem{example}{\hspace{-11pt}\bf Example}

\newtheorem{remark}{\hspace{-1pt}\bf Remark}

\begin{document}

\title{ Underlay Cognitive Radios with Capacity Guarantees for Primary Users}

\author{
{ Antonio G. Marques}\\
\thanks {The work in this
paper was supported by the Spanish Min. of Sci.\&Inn. grant No.
TEC2009-12098. Parts of this paper were presented at CROWNCOM 2012. This paper has been submitted for publication to IEEE TSP.}
\thanks{A. G. Marques is with the Dept. of Signal Theory and Comms., King Juan Carlos Univ., Camino del Molino s/n, Fuenlabrada, Madrid 28943,
Spain. Phone: +34 914-888-222, fax: +34 914-887-500,
email:~antonio.garcia.marques{\rm\char64}urjc.es}
}
\vspace{-1.8cm}


\vspace{-1.8cm}
\maketitle
\vspace{-1.8cm}
\begin{abstract}
To use the spectrum efficiently, cognitive radios leverage knowledge of the channel state information (CSI) to optimize the performance of the secondary users (SUs) while limiting the interference to the primary users (PUs). The algorithms in this paper are designed to maximize the weighted ergodic sum-capacity of SUs, which transmit orthogonally and adhere simultaneously to constraints limiting: i) the long-term (ergodic) capacity loss caused to each PU receiver; ii) the long-term interference power at each PU receiver; and iii) the long-term power at each SU transmitter. Formulations accounting for short-term counterparts of i) and ii) are also discussed. Although the long-term capacity constraints are non-convex, the resultant optimization problem exhibits zero-duality gap and can be efficiently solved in the dual domain. The optimal allocation schemes (power and rate loadings, frequency bands to be accessed, and SU links to be activated) are a function of the CSI of the primary and secondary networks as well as the Lagrange multipliers associated with the long-term constraints. The optimal resource allocation algorithms are first designed under the assumption that the CSI is perfect, then the modifications needed to accommodate different forms of imperfect CSI (quantized, noisy, and outdated) are analyzed.
\end{abstract}
\begin{keywords}
Cognitive radios, resource management, stochastic approximation, imperfect channel state information.
\end{keywords}

\section{Introduction}\label{S:Introduction}

Cognitive radios (CRs) implementing dynamic spectrum access (DSA) schemes are the next generation solution for the problem of deploying new wireless services in an overcrowded radio environment \cite{Haykin05,GoldsmithCR09}.
CR users, typically referred to as secondary users (SUs), have to sense the radio spectrum and use the sensing measurements to adapt dynamically the configuration of the CR. Such tasks have to be carried out with the aim of optimizing the quality of service (QoS) of the SUs while limiting the interference to the receivers which hold the licence of the frequency band, referred to as primary users (PUs). The specific rules that establish how SUs and PUs coexist and how the interference is limited depend on the so-called CR paradigm considered (underlay, overlay, or interweave \cite{GoldsmithCR09}) and the DSA policy implemented \cite{SurvZhaoSadler07}.

The merits of adaptive schemes for traditional wireless systems that first acquire knowledge of the channel state information (CSI) and then use the CSI to optimally allocate the transmit resources are well documented; see \cite{GoldsmithBook}. However, for channel-adaptive schemes to be deployed in CR scenarios \cite{refInterf2,NeelyCR09,Joint_S_RA_Xin10,RA_CR_Kang09}, important \emph{challenges} not present in traditional wireless networks arise. Next we describe several of them.

\emph{Challenge 1}: Sensing the CR spectrum and acquiring the corresponding CSI (especially the one of the primary network) is a difficult task. The CSI in CRs is heterogeneous (presence of PUs, SU-to-PU channels, SU-to-SU channels, PU-to-PU channels) and inherently distributed. Some PUs can be located far away and not willing to collaborate with the SUs. The CSI may also vary fast and, due to interference, might not be stationary. Furthermore, to become aware of the \emph{overall radio environment}, not only channels but also additional (network) side information may need to be sensed/estimated \cite{GoldsmithCR09}. As a result, the CSI in CRs has higher dimensionality and heterogeneous quality (information of SU-to-SU links is typically better than that of SU-to-PU). Hence, advanced signal processing schemes that keep track of the CSI and mitigate the existing uncertainties have to be implemented. To deal with these problems, most CR works consider that the CSI contains some type of imperfections. Such imperfections are typically modeled as either noisy CSI (the actual CSI is corrupted with additive noise \cite{refInterf2}) or quantized CSI (only a coarse description of the channel CSI is available, \cite{quantizedCR,SI_jsac_lrf}). Fewer works have considered the fact that the CSI may be not only noisy but also outdated \cite{QZhao08,JSAC}; have developed signal processing schemes to mitigate the CSI uncertainties; or have incorporated those imperfections into the design of resource allocation (RA) algorithms \cite{refInterf2,Imperf_CRref2,Imperf_CRref4,emiliano11twc}. In this paper we take a general approach to model the CSI imperfections and consider that the distribution of the instantaneous CSI (referred to as belief) is available. This will allow us to: i) consider simultaneously different sources of CSI imperfections; and ii) address the design of systems with a broad degree of CSI uncertainties (from almost perfect CSI to severely degraded CSI). The expression for the belief and the rules to update it will depend on the operating conditions of the system. For example, if the CSI is perfect, the belief coincides with the instantaneous channel measurements. On the other hand, if only statistical CSI is available, the belief coincides with the long-term distribution of the channel and does not vary with time.

\emph{Challenge 2}: As already mentioned, CR transmissions must obey additional rules that establish how SUs and PUs coexist and how to control interference. Such rules are typically formulated as constraints and depend on the specific CR paradigm and the DSA policies implemented. Overlay CRs (referred to as interweave CRs in \cite{GoldsmithCR09}) allow SUs to transmit only if PUs are not active. Differently, underlay CRs allow for SU transmissions provided that the damage (interference) to the PUs is not too high. To keep the interference low, some works limit the interference power at the primary receiver side, either by imposing instantaneous (short-term) or average (long-term) interference power constraints; see, e.g., \cite{RA_CR_Kang09,CognitiveDiversity09,SurveyConvexCR_underlay_SPMag10,Sergiy_icassp11}. The latter are better suited for fading channels because they can exploit the diversity of the interfering link \cite{CognitiveDiversity09,Sergiy_icassp11}. Other works guarantee a minimum signal-to-interference-plus-noise ratio (SINR) at the PU receiver \cite{SINR1,GatsisDSA}. Short-term SINR constraints can be easily translated to (short-term) interference power constraints, while long-term SINR constraints cannot. More recent designs use a probabilistic approach to limit the probability of interfering the primary transmissions \cite{NeelyCR09,Joint_S_RA_Xin10,Alouini11,JSAC}. Other works have designed schemes either guaranteeing a minimum capacity (rate) for the PU or limiting the capacity-loss at the PU receiver \cite{GatsisDSA,quantizedCR}. Providing guarantees on the capacity of the PU links is typically a non-convex problem, so that most works have developed suboptimal solutions and focused on short-term formulations, which are more tractable and in some cases can be rendered convex \cite{GatsisDSA}. In this paper we consider that PUs are not always active. When the channels are not occupied, the SUs are allowed to transmit (overlay paradigm). When the PUs are active, the SUs transmissions adhere to diverse DSA constraints (short and long term interference power and rate loss) that guarantee that the damage to PUs is kept under control (underlay paradigm).

\emph{Challenge 3}: CRs have to use the time-varying (imperfect) CSI to dynamically adapt the available resources (power and rate loadings of the SUs) and decide the frequency bands to be used and the specific SUs that will use them. Relative to the RA in traditional wireless systems, the problem in CRs is challenging not only because more variables are involved, but also because the description of the CSI is more complicated and the schemes have to satisfy the additional DSA constraints. Different approaches have been used to formulate and solve the RA problem: game theory \cite{CR_jsac_GameTheory}, non-linear optimization \cite{SurveyConvexCR_underlay_SPMag10}, convex approximation \cite{emiliano11twc}, dynamic programming \cite{QZhao08}, adaptive control \cite{NeelyCR09} and even bio-inspired models \cite{DiLorenzoBarbarossa_tsp11}. In this paper, we design the RA schemes using non-linear optimization and dual stochastic approximation tools. The stochastic schemes are robust to channel non-stationarities and require less computational burden than that of the (non-stochastic) allocation schemes. Moreover, they are well suited for dealing with CSI imperfections. Dual stochastic algorithms have been successfully used to allocate resources in wireless networks, see, e.g., \cite{AleTSP11,am_etal_tvt12} and \cite{quantizedCR,Joint_S_RA_Xin10} for examples in the context of CRs.

Motivated by these challenges, we design RA algorithms that optimize the rate performance of the SUs and limit the interference to the PUs. We focus on CRs where SUs adapt their power and rate loadings dynamically, and access orthogonally a set of frequency bands which are primarily devoted to PU transmissions. Orthogonal here means that if a SU is transmitting, no other SU can be active in the same band. The RA schemes are then obtained as the solution of a weighted sum-average capacity maximization subject to four types of constraints: i) limits on the long-term (ergodic) capacity loss inflicted to each PU; ii) limits on the long-term interference power at each PU \cite{Sergiy_icassp11}; iii) limits on the long-term power transmitted by each SU; and iv) short-term formulations of i) and ii). Consideration of i) is challenging because the interfering (SU) powers render the capacity term non-convex, and it is the main contribution of this work. Although non-convex, it holds that the formulated problem has zero duality gap. As a result, the Langrangian relaxation is optimal. Additionally, the operating conditions of the secondary network (and the formulation of the objective to optimized) are such that the problem in the dual domain can be separated across users and frequency bands. This favorable structure allows for a significant reduction on the complexity required to find the optimal solution and, hence, renders the non-convex problem computationally tractable. Different forms of channel imperfections are considered (quantized, noisy, outdated, statistical). The optimal RA schemes are complemented with simple but effective stochastic signal processing algorithms both to mitigate the effects of the CSI imperfections, and to estimate online the value of the multipliers required to implement the optimal RA. Such stochastic algorithms are able to track the time-variation of the environment and/or learn unknown parameters on-the-fly, features that are especially attractive for CR systems \cite{Haykin05,quantizedCR}.

The rest of the paper is organized as follows. Sec. II presents the model for the (perfect) CSI, describes the operating conditions of the secondary network, and formulates the DSA constraints that SUs must obey. Sec. III deals with the design of the optimal RA algorithms. First, the optimization problem which gives rise to the RA is formulated and then, its solution is obtained. Sec. IV discusses different methods (including stochastic) to estimate the multipliers required to implement the optimal RA. Sec. V describes different forms of CSI imperfections and analyzes how the optimal schemes have to be modified to account for imperfect CSI. Sec. VI presents different illustrative numerical examples that corroborate the theoretical claims. Conclusions in Sec. VII wrap-up this paper.\footnote{ {\emph Notation:} $^T$ denotes vector transposition; $x^*$ the optimal value of variable $x$; $\wedge$ ($\vee$) the Boolean ``and'' (``or'') operator; $\mathbbm{E}[\cdot]$ expectation; $\mathbbm{1}_{\{\cdot\}}$
the indicator function ($\mathbbm{1}_{\{x\}}=1$ if $x$ is true and
zero otherwise); and $[x]_a^b$ the projection of the scalar $x$ onto the interval $[a,b]$, i.e., $[x]_a^b:=\min\{\max\{x,a\},b\}$. }\vspace{-0.2cm}

\section{Model description}\label{S:Preliminaries-model}

We consider a CR network with $M$ secondary users (indexed by $m$) transmitting opportunistically and orthogonally over $K$ different frequency bands (indexed by $k$). For simplicity, we assume that: i) each band has the same bandwidth and is occupied by a different primary user; and ii) the secondary network has an access point (AP) which is the destination of all secondary users. The AP acts as a central scheduler which collects the CSI and then makes the RA decisions. Extensions to scenarios where those assumptions do not hold true can be handled with a moderate increase in complexity.

\subsection{Channel state information}

Intuitively speaking, the CSI in wireless systems comprises the information of the channel links which: i) is known by the system and ii) is relevant from a RA perspective. A key feature of CR systems is that the CSI is heterogeneous, meaning that it is typically different for the primary and secondary network. The reason for that is twofold. First, the schemes used to acquire the CSI are different for the primary and secondary network [cf. i)]. Second, the impact of the CSI on the design of the RA is different [cf. ii)]. For ease of exposition, we first design the RA schemes assuming that the CSI is error-free. Accordingly, the model for the perfect CSI is presented here, while the model for imperfect CSI (and the corresponding modifications for the RA schemes) is presented in Sec. \ref{S:ImperfectCSI}.

The CSI available at instant $n$ is formed by variables: $a_{k,1}[n]$, $h_{k,1}^m[n]$, and $h_{k,2}^m[n]$ for all $k$ and $m$. Before explaining the meaning of such variables, we clarify that subscript ``1'' will be used to emphasize that the channel involves \emph{primary} transceivers, while subscript ``2'' is used to emphasize that only \emph{secondary} transceivers are involved. Starting with the CSI of the PUs, $a_{k,1}[n]$ is a Boolean variable which is one if the PU that transmits on the $k$th channel is \emph{active} at time $n$ and zero otherwise. Variable $h_{k,1}^m[n]$ represents the instantaneous noise-normalized power gain between the $m$th SU and the $k$th PU at instant $n$. Similarly, $h_{k,2}^m[n]$ represents the instantaneous noise-normalized power gain between the $m$th SU and the AP in the $k$th channel at instant $n$. All $a_{k,1}[n]$, $h_{k,1}^m[n]$ and $h_{k,2}^m[n]$ are stationary random processes. The assumption of perfect CSI implies that at instant $n$, the value of those variables is known deterministically. Finally, we will use $\gamma_k$ to denote the (interference free) signal-to-noise ratio (SNR) between the PU transmitter and PU receiver. For simplicity, we will assume that $\gamma_k$ does not vary with time (either because the PU channels are fixed or because the PU transmitter implements a channel-inversion power loading \cite{GoldsmithBook}). Nonetheless, our schemes can be easily modified to account for $\gamma_k$ varying with time.

To finish this section, let $\mathbf{h}$ denote the $K(2M+1)\times 1$ vector of overall CSI containing: i) the power gains of the $MK$ CR-to-CR links, and ii) the normalized power gains of the $MK$ CR-to-PU links; and iii) $K$ Boolean variables indicating whether the channels are occupied. Clearly, the value of $\mathbf{h}$ varies with time and, wherever convenient, we will write $\mathbf{h}[n]$ to stress this fact.

\subsection{Resources at the secondary network}\label{Ss:ResSecNet}
Now, we introduce the design variables, i.e, the variables that will be adapted as a function of the (primary and secondary) CSI $\mathbf{h}$. Let $w_{k,2}^m$ denote a Boolean variable taking the value one if the $m$th \emph{secondary} user is scheduled to transmit into the $k$th band and zero otherwise. Provided that $w_{k,2}^m=1$, let $p_{k,2}^m$ denote the instantaneous power transmitted over the $k$th band
 by the $m$th \emph{secondary} user. We analyze the case where instantaneous rate and power variables are coupled through Shannon's capacity formula. Such a coupling will be written as $r_{k,2}^m(h_{k,2}^m p_{k,2}^m):=\log_2(1+h_{k,2}^m p_{k,2}^m)$, which is an increasing and concave function. Nonetheless, the basic results in this paper hold for any $r_{k,2}^m(\cdot)$ increasing and concave.

The CR operates in a time-block fashion, where the duration of each block corresponds to the coherence time of the fading channel. This way, at every time $n$ the AP will use the current CSI vector
$\mathbf{h}$ to find the (optimum) value of $w_{k,2}^m$ and $p_{k,2}^m$. Since $\mathbf{h}$ varies with $n$ and $\{w_{k,2}^m, p_{k,2}^m\}$ depend on $\mathbf{h}$, the value of the design variables $\{w_{k,2}^m, p_{k,2}^m\}$ will vary across time as well. Throughout the manuscript, we will write $\mathbf{h}$, $w_{k,2}^m(\mathbf{h})$ and $p_{k,2}^m(\mathbf{h})$, or $\mathbf{h}[n]$, $w_{k,2}^m[n]$ and $p_{k,2}^m[n]$, wherever is convenient to emphasize the corresponding dependence.

Having introduced the design variables, now we formulate constraints that these variables need to satisfy. To ensure that at most one user transmits into a given band $k$, we need $\sum\nolimits_{m}w_{k,2}^m(\mathbf{h}) \leq 1$. If the left hand side of the constraint is equal to one, then one user is accessing the channel (orthogonal access). If it is equal to zero, then none is transmitting (either because all secondary channels are poor, or because it causes very high interference to the PUs). To simplify the notation, we consider an additional virtual SU user $m=0$, with zero transmit power and rate; i.e., $p_{k,2}^0=r_{k,2}^0=0$. The $0$th user will be active (and thus $w_{k,2}^0=1$) if none of the actual SUs is transmitting. Then, we can write
\begin{equation}\label{E:optRA_c_inst_sched}
\sum\nolimits_{m}w_{k,2}^m(\mathbf{h}) = 1,~~~\forall k.
\end{equation}
We also consider that the maximum average (long-term) power the $m$th SU can transmit is  $\check{p}_2^m$; hence,
\begin{equation}\label{E:optRA_c_av_pow}
\mathbbm{E}_{\mathbf{h}}\left[\sum\nolimits_{ k}w_{k,2}^m(\mathbf{h})p_{k,2}^m(\mathbf{h})\right]\leq\check{p}_2^m,~~\forall m.
\end{equation}
Such a constraint is not only reasonable to effect QoS across CRs, but also to limit the power consumption of each of the CR transmitters. The expectation in \eqref{E:optRA_c_av_pow} is taken over all possible values of $a_{k,1}[n]$ $h_{k,1}^m[n]$ and $h_{k,2}^m[n]$; i.e, considering all $m$, $k$, and $n$. While \eqref{E:optRA_c_inst_sched} needs to hold for each and every channel realization (hence, for each and every time instant), \eqref{E:optRA_c_av_pow} only needs to hold in the long term.

\subsection{Dynamic spectrum access constraints}\label{Ss:CRConstraints}
The next step is to identify the rules that dictate how SU transmissions affect the performance of the PUs. Such rules will be formulated as constraints that will be incorporated into the optimization problem that gives rise to the RA schemes. In other words, the DSA constraints will represent how SUs have to modify their behavior so that the damage caused to the PUs is kept under control.

When the DSA constraints are formulated, several factors have a significant impact both in terms of the system operation and the mathematical formulation of the problem. Two important ones are discussed next. The first factor is whether the interference constraints are formulated as instantaneous (short-term) or as average (long-term) constraints. The former requires the constraint to hold for each and \emph{every time instant}, while the latter requires the constraint to hold on average (taking into account all time instants jointly). Clearly, instantaneous constraints are more restrictive than their average counterparts (which can exploit the so-called ``cognitive diversity'' of the primary CSI \cite{CognitiveDiversity09,SurveyConvexCR_underlay_SPMag10}), and therefore the performance of the secondary network will be higher in the latter case. Mathematically, long-term constraints are typically dualized, while short-term constraints are handled using alternative methods. The second factor is the metric used to measure the actual damage that the CRs inflict to the PUs. Among the metrics considered in the literature we find: interference power at the PUs, probability on interfering the PUs, and rate loss inflicted to the PUs. Most works have focused on limiting the interference power. The reason is twofold: i) it is a simple (and intuitive) metric to measure the interference, and ii) it can be formulated as a convex constraint. Limiting the rate loss may be considered a better alternative because it focuses on the actual damage that the interference causes to the PUs (most communications systems are designed to either guarantee or maximize a certain transmission rate). From a mathematical perspective, constraints limiting the rate loss are typically non-convex. As a result, very few works have explored that alternative; see e.g. \cite{GatsisDSA,quantizedCR}. The problem of limiting the probability of interference for a system with operating conditions very similar to the ones considered in this paper was thoroughly investigated in \cite{JSAC}.

As already mentioned, the main contribution of this work is to limit the long-term rate (capacity) loss on the PUs. However, we will also impose limits on the long-term interference power. The reason is twofold. First, such constraints were not considered for systems with the \emph{same exact} operating conditions than those considered in this work; see \cite{Sergiy_icassp11} for a very related one. More importantly, joint consideration of rate loss and interference power constraints will help us to compare these two alternatives. For similar reasons, the end of the section is devoted to discuss the modifications required to handle \emph{short-term} interference power and rate loss constraints.

We start with the formulation of the long-term \emph{interference power} constraints. Let $\check{p}_{k,1}$ denote the maximum average interference power the $k$th primary receiver can tolerate (provided that the PU is active) and recall that the $m$th SU transmits in the $k$th channel only if the Boolean scheduling variable $w_{k,2}^m(\mathbf{h})$ is one. Then, the following $K$ constraints need to hold
\begin{equation}\label{E:optRA_c_av_power_interf}
\mathbbm{E}_{\mathbf{h}}\left[\sum_m w_{k,2}^m(\mathbf{h}) h_{k,1}^m p_{k,2}^m(\mathbf{h})~\Big|a_{k,1}=1\right]\leq \check{p}_{k,1},~~\forall k.
\end{equation}
The fact that the expectation is taken across all $\mathbf{h}$ reflects that \eqref{E:optRA_c_av_power_interf} is a long-term constraint. Clearly, for a given channel realization $\mathbf{h}$ just one of the $M+1$ terms inside the expectation is active. This property will be exploited in upcoming sections. Finally, note that only CSI realizations for which $a_{k,1}=1$ are considered in the expectation. In fact, \eqref{E:optRA_c_av_power_interf} can be rewritten as $\mathbbm{E}_{\mathbf{h}}[a_{k,1}\sum_m w_{k,2}^m(\mathbf{h}) h_{k,1}^m p_{k,2}^m(\mathbf{h})]\leq \mathbbm{E}_{\mathbf{h}}[a_{k,1}\check{p}_{k,1}]$. If one does not want to bound the long-term interference power that the PU receives when it is active, but the long-term power at the PU receiver irrespective of whether the PU is active of not, then $a_{k,1}$ has to be removed from the previous expressions.

Next, we formulate the long-term (ergodic) capacity constraints. For such a purpose we define the function $r_{k,1}(x):=\log_2\left(1+\frac{\gamma_{k,1}}{1+x}\right)$, where $x$ stands for the interference power at the $k$th PU receiver. Our formulation guarantees a minimum long-term rate of $\check{r}_{k,1}$ for the $k$th PU. This minimum rate can either be a fixed value \cite{quantizedCR} or expressed as a fraction of the rate that the PU achieves when no CRs are present. Mathematically, the rate requirement in the latter case can be written as $\check{r}_{k,1}:=(1-\check{\varepsilon}_k)\mathbbm{E}_{\mathbf{h}}\left[ a_{k,1}r_{k,1}(0)\right]$ where $\check{\varepsilon}_k\in(0,1)$ is the maximum (relative) capacity loss that the CRs can cause to the $k$th PU. With these issues in mind, the long-term \emph{capacity constraint} is formulated as
\begin{equation}\label{E:optRA_c_av_rate_loss}
\mathbbm{E}_{\mathbf{h}}\left[\sum_m w_{k,2}^m(\mathbf{h}) r_{k,1}(h_{k,1}^m p_{k,2}^m(\mathbf{h}))\hspace{0.05cm}\Big|a_{k,1}\hspace{-0.05cm}=\hspace{-0.05cm}1\right]\geq \check{r}_{k,1},~\forall k.
\end{equation}
Again, for a given channel realization $\mathbf{h}$ only one of the $M+1$ terms inside the expectation is active. The expression in \eqref{E:optRA_c_av_rate_loss} confirms that if the constraint is written as $f(p_{k,2}^m(\mathbf{h}))\leq0$, then $f(\cdot)$ is a non-convex function [cf. the definition of $r_{k,1}(\cdot)$].

We close this section by briefly discussing the formulation of the short-term DSA constraints. To write the short-term counterparts of \eqref{E:optRA_c_av_power_interf} and \eqref{E:optRA_c_av_rate_loss} we do not need to take into account all $\mathbf{h}$, but only the current one $\mathbf{h}[n]$. Hence, the short-term constraints for the time instant $n$ are
\begin{eqnarray}\label{E:optRA_c_inst_power_interf}
a_{k,1}[n]\sum_m w_{k,2}^m[n] h_{k,1}^m[n] p_{k,2}^m[n]\leq a_{k,1}[n]\check{p}_{k,1},\\
\label{E:optRA_c_inst_rate_loss}
a_{k,1}[n]\sum_m w_{k,2}^m[n] r_{k,1}(h_{k,1}^m[n] p_{k,2}^m[n])\geq a_{k,1}[n]\check{r}_{k,1},
\end{eqnarray}
which need to hold for all $k$ and $n$. Capitalizing on the fact that at every time instant only one SU is active, the alternative set of constraints can be considered
\begin{eqnarray}
\label{E:optRA_c_inst_power_interf_ind_users}
a_{k,1}[n]h_{k,1}^m[n]p_{k,2}^m[n]\leq a_{k,1}[n]\check{p}_{k,1},\\
\label{E:optRA_c_inst_rate_loss_ind_users}
a_{k,1}[n]r_{k,1}(h_{k,1}^m[n] p_{k,2}^m[n])\geq a_{k,1}[n]\check{r}_{k,1},
\end{eqnarray}
which in this case need to hold for all $k$, $m$ and $n$. Clearly, if \eqref{E:optRA_c_inst_power_interf_ind_users} and \eqref{E:optRA_c_inst_rate_loss_ind_users} are satisfied, then \eqref{E:optRA_c_inst_power_interf} and \eqref{E:optRA_c_inst_rate_loss} are satisfied too. It can also be rigorously shown that \eqref{E:optRA_c_inst_power_interf_ind_users} and \eqref{E:optRA_c_inst_rate_loss_ind_users} do not imply a loss of optimality relative to \eqref{E:optRA_c_inst_power_interf} and \eqref{E:optRA_c_inst_rate_loss}. As already pointed out, key for showing this result is that at every time instant at most one SU is active, so that bounds on the non-active users are irrelevant. The main advantage of considering \eqref{E:optRA_c_inst_power_interf_ind_users} and \eqref{E:optRA_c_inst_rate_loss_ind_users} is that the transmit powers of the different SUs are decoupled, so that each of the $MK$ expressions in \eqref{E:optRA_c_inst_power_interf_ind_users} and \eqref{E:optRA_c_inst_rate_loss_ind_users} can be solved with respect to (w.r.t.) $p_{k,2}^m[n]$. This implies that the constraints can be rewritten as simple box constraints. To be specific, let $\check{p}_{k,\max}^m$ represent the maximum power the amplifier at the SU can transmit. Moreover, assume that $a_{k,1}[n]=1$ and let $x_k^m[n]$ and $y_k^m[n]$ be, respectively, the values of $p_{k,2}^m[n]$ for which the constraints \eqref{E:optRA_c_inst_power_interf_ind_users} and \eqref{E:optRA_c_inst_rate_loss_ind_users} are satisfied with equality. Based on these notational conventions, we define the maximum short-term power as $\check{p}_{k,2}^m[n]:=\check{p}_{k,\max}^m$ if $a_{k,1}[n]=0$, and $\check{p}_{k,2}^m[n]:=\min\{x_k^m[n],y_k^m[n],\check{p}_{k,\max}^m\}$ if $a_{k,1}[n]=1$. Then, the short-term DSA constraints can be replaced with $p_{k,2}^m[n]\leq \check{p}_{k,2}^m[n]$. In a nutshell, the orthogonal access among SUs allow us to rewrite the short-term DSA constraints as \emph{time-varying} power \emph{peak} constraints. The power bound enforced by each of such peak constraints will depend on the metrics used to measure the interference (rate loss and/or interference power), the limits set on the chosen metric ($\check{p}_{k,1}$ and $\check{r}_{k,1}$), and the CSI at instant $n$.

\section{Formulating and solving the RA problem}\label{S:ProblemFormulation}
\vspace{-.05cm}

To formulate the optimization problem that gives rise to the optimum RA algorithms, we need to identify: i) the variables to be optimized; ii) the constraints the variables need to satisfy; and iii) the metric to be optimized. The first step was accomplished in Sec. \ref{Ss:ResSecNet}. Regarding the second step, Boolean variables $w_{k,2}^m(\mathbf{h})$ are constrained to belong to the set $\{0,1\}$ and variables $p_{k,2}^m(\mathbf{h})$ are constrained to belong to the set $[0,\check{p}_{k,2}^m(\mathbf{h})]$, where $\check{p}_{k,2}^m(\mathbf{h})$ stands for the instantaneous peak power constraint introduced at the end of Sec. \ref{Ss:CRConstraints}. Moreover, $w_{k,2}^m(\mathbf{h})$ and $p_{k,2}^m(\mathbf{h})$ need to satisfy \eqref{E:optRA_c_inst_sched} and \eqref{E:optRA_c_av_pow}, and the DSA constraints in \eqref{E:optRA_c_av_power_interf} and \eqref{E:optRA_c_av_rate_loss}.

Regarding the third step (metric to be optimized), we are interested in maximizing the weighted ergodic sum-capacity given by $\bar{c}_2:=\sum\nolimits_{ k,m}\mathbbm{E}_{\mathbf{h}}\left[\beta^m w_{k,2}^m(\mathbf{h})r_{k,2}^m(h_{k,2}^m p_{k,2}^m(\mathbf{h}))\right]$, where $\beta^m>0$ represents a user-dependent priority coefficient. Note that by varying $\{\beta^m\}_{m=1}^M$, the border of the capacity region can be found \cite{XinInfoTheory}. Recall that for a given channel realization $\mathbf{h}$ and channel $k$ only one of the $M+1$ terms (SUs) is active. Other objective functions, such as ergodic sum-utility rate could be used without changing the basic structure of the solution; see, e.g., \cite{am_etal_tvt12} for further details on a related problem.

Under all previous considerations, the optimal RA is obtained as the solution of the following problem:
\vspace{-.1cm}
\begin{subequations}
\label{E:optRA}
\begin{alignat}{2}
\label{E:optRA_obj}
        \hspace{-0.03cm}\bar{c}_2^*:=\underset{\{w_{k,2}^m(\mathbf{h}),p_{k,2}^m(\mathbf{h})\}}{\max}
       ~{\displaystyle \sum\nolimits_{ k,m}\hspace{-0.13cm}\mathbbm{E}_{\mathbf{h}}\left[\beta^m w_{k,2}^m(\mathbf{h})r_{k,2}^m(h_{k,2}^mp_{k,2}^m(\mathbf{h}))\right]}\hspace{0.06cm}\\
\label{E:optRA_inst_const}\hspace{-1.3cm}~\mathrm{s.~to:}~~w_{k,2}^m(\mathbf{h})\in \{0,1\}, ~ 0\leq p_{k,2}^m(\mathbf{h})\leq \check{p}_{k,2}^m(\mathbf{h}),~\eqref{E:optRA_c_inst_sched};\hspace{1.0cm}\\
\label{E:optRA_av_const}
\hspace{-1.3cm}~~\eqref{E:optRA_c_av_pow},~~\eqref{E:optRA_c_av_power_interf},~~\eqref{E:optRA_c_av_rate_loss};\hspace{5.8cm} \end{alignat}
\end{subequations}
where the dependence of the optimization variables on the CSI $\mathbf{h}$ has been made explicit. Note that we are interested in optimizing a long-term objective \eqref{E:optRA_obj}, subject to both short-term \eqref{E:optRA_inst_const} and long-term \eqref{E:optRA_av_const} constraints. As we will see in the next section, the approach to handle \eqref{E:optRA_inst_const} and \eqref{E:optRA_av_const} will not be the same.

\subsection{Optimal RA}

The main challenge of finding the optimal RA is that \eqref{E:optRA} is not a convex problem. Basically, there are three sources of non-convexity in \eqref{E:optRA}: i) scheduling coefficients $w_{k,2}^m$ are constrained to belong to $\{0,1\}$, which is a non-convex set; ii) the monomials $w_{k,2}^m p_{k,2}^m$, and $w_{k,2}^m r_k^m$ are not jointly convex; and iii) the constraint \eqref{E:optRA_c_av_rate_loss} is not convex w.r.t. $p_{k,2}^m$. The two first sources on non-convexity can be ``easily'' bypassed by transforming (relaxing) the problem in \eqref{E:optRA} into a convex one which yields the same optimality conditions; see App. A for technical details. However, the third source of non-convexity cannot be bypassed. Two undesirable consequences associated with lack of convexity are \cite{Bertsekas_Book_CA03}: (c1) zero-duality gap is not guaranteed, and (c2) development of numerical algorithms that find the optimal solution in polynomial time is not guaranteed. Remarkably, it can be shown that (see related discussion in App. A, and \cite{AleGG10ZeroDualityGap}, \cite{KetanNikosZeroDualGap}): \emph{the problem in \eqref{E:optRA} exhibits zero-duality gap}. This result implies that the constraints can be dualized without losing optimality. However, (c2) still holds, so that finding an efficient algorithm to optimize the (unconstrained) Lagrangian is still challenging. Interestingly, due to the structure of \eqref{E:optRA} we will show that the optimization can be separated (decomposed) across channels and users, decreasing dramatically the computational complexity to find the optimal solution.

After the previous discussion, we are ready to present the solution of \eqref{E:optRA}. Our approach to deal with the constraints in \eqref{E:optRA} is twofold. The long-term constraints in \eqref{E:optRA_av_const} --namely, \eqref{E:optRA_c_av_pow}, \eqref{E:optRA_c_av_power_interf} and \eqref{E:optRA_c_av_rate_loss}-- will be dualized, while the constraints in \eqref{E:optRA_inst_const} (all short-term) will be handled using alternative methods such as scalar projections. Regarding the long-term constraints, let $\pi^m$, $\theta_k$ and $\rho_k$ denote the Lagrange multipliers associated with \eqref{E:optRA_c_av_pow}, \eqref{E:optRA_c_av_power_interf} and \eqref{E:optRA_c_av_rate_loss}, respectively. With this notational conventions, it can be shown (see App. A) that the optimal solution of \eqref{E:optRA} is
\vspace{-.2cm}
\begin{eqnarray}
\nonumber\varphi_k^m(p_{k,2}^m[n])\hspace{-.3cm}&:=&\hspace{-.3cm}\beta^m r_{k,2}^m(h_{k,2}^m[n]p_{k,2}^m[n])-\pi^mp_{k,2}^m[n]\\
\nonumber\hspace{-.3cm}&-&\hspace{-.3cm}\theta_k a_{k,1}[n]h_{k,1}^m[n]p_{k,2}^m[n]\\
\label{E:opt_ind_case1}\hspace{-.3cm}&+&\hspace{-.3cm}\rho_k a_{k,1}[n]r_{k,1}(h_{k,1}^m[n] p_{k,2}^m[n]),\hspace{.3cm}\\
\label{E:opt_pow_case1}p_{k,2}^{m*}[n]\hspace{-.3cm}&:=&\hspace{-.3cm}\Big[\arg\max_{p_{k,2}^m[n]}~\varphi_k^m(p_{k,2}^m[n])\Big]_0^{\check{p}_{k,2}^m[n]},\\
\label{E:opt_sched_case1}w_{k,2}^{m*}[n]\hspace{-.3cm}&:=&\hspace{-.3cm}\mathbbm{1}_{\{ m=\arg\max_{l} \varphi_k^l(p_k^{l*}[n])\}}\mathbbm{1}_{\{p_{k,2}^{m*}[n]>0\hspace{.05cm}\vee\hspace{.05cm}m=0\}}\hspace{-.06cm}.\hspace{.3cm}
\end{eqnarray}
Key for understanding the solution of \eqref{E:optRA} is the definition of the functional $\varphi_k^m(\cdot)$ in \eqref{E:opt_ind_case1}. Mathematically, $\varphi_k^m(x)$ represents the contribution to the \emph{Lagrangian} of \eqref{E:optRA} if the transmit power is $p_{k,2}^m[n]=x$ and $w_{k,2}^m[n]=1$. Intuitively, \eqref{E:opt_ind_case1} can be interpreted as a user-channel quality indicator (the higher the indicator, the better). Under this interpretation, the rates of SUs and PUs are rewards (first and fourth terms), and the transmit and interference powers are costs (second and third terms). The corresponding prices are $\beta^m$, $\rho_k$, $\pi^m$ and $\theta_k$, respectively. The indicator also manifests the existing trade-off between the SUs (first and second terms) and the PUs (third and forth terms). Note that if the fourth term in \eqref{E:opt_ind_case1} is replaced with $-\rho_k a_{k,1}[n]\Big(r_{k,1}(0)-r_{k,1}(h_{k,1}^m[n] p_{k,2}^m[n])\Big)$, the optimum value of $p_{k,2}^{m*}[n]$ and $w_{k,2}^{m*}[n]$ in \eqref{E:opt_pow_case1} and \eqref{E:opt_sched_case1} do not change. This implies that we can also interpret the quality indicator as a functional which penalizes the allocations that entail a high capacity loss for the PU.

Based on the definition $\varphi_k^m(p_{k,2}^m[n])$, equation \eqref{E:opt_pow_case1} reveals that $p_{k,2}^{m*}[n]$ is found separately for each of the user-channel pairs. Similarly, \eqref{E:opt_sched_case1} reveals that to find $\{w_{k,2}^{m*}[n]\}_{m=0}^M$, i.e., the optimal scheduling for channel $k$; no information from channels other than $k$ is required. These attractive features are present because the optimization problem in the dual domain is separable across users and channels (see \cite{am_etal_tvt12}, \cite{JSAC}). Keys for this property to hold are the consideration of orthogonal access in the secondary network and the definition of the objective in \eqref{E:optRA}.
Capitalizing on the favorable structure of the solution, we now analyze in further detail the optimal RA. Starting with the optimal scheduling in \eqref{E:opt_sched_case1}, we observe that $w_{k,2}^{m*}[n]$ is available in closed form, provided that the optimum power is known. Equation \eqref{E:opt_sched_case1} reveals that the scheduling follows a winner-takes-all strategy, guaranteeing that the access is orthogonal (at most one user is active), opportunistic ($\varphi_k^m$ is a continuous random variable), and greedy (only the user with \emph{highest} quality in a given band must be scheduled). Note that the second condition in \eqref{E:opt_sched_case1} dictates that if all users decide to transmit with zero power, the channel is assigned to the virtual user $m=0$. The details of the optimum power allocation are a bit more intricate. To obtain $p_{k,2}^{m*}[n]$ we need first to maximize $\varphi_k^m(p_{k,2}^m[n])$ w.r.t. $p_{k,2}^m[n]$. Consider first a simplified case where the CR constraints \eqref{E:optRA_c_av_power_interf} and \eqref{E:optRA_c_av_rate_loss} are not present. In such a case only the two first terms in \eqref{E:opt_ind_case1} are present, so that $\varphi_k^m(\cdot)$ is strictly concave and differentiable. As a result, the optimization is convex and $p_{k,2}^{m*}[n]$ can be easily found. Specifically, $p_{k,2}^{m*}[n]$ for this case is available in closed form as $p_{k,2}^{m*}[n]=[\frac{\beta^m\log_2(\exp(1))}{\pi_m}-\frac{1}{h_{k,2}^m}]_0^{\check{p}_{k,2}^m[n]}$. The previous expression is basically a waterfilling power loading \cite{GoldsmithBook} projected onto the feasible interval defined by the instantaneous constraints. When the CR constraint \eqref{E:optRA_c_av_power_interf} is active, the third term in \eqref{E:opt_ind_case1} needs to be considered. However, since that term is linear w.r.t. $p_{k,2}^m[n]$, the structure of $\varphi_k^m(\cdot)$ is basically the same and $p_{k,2}^{m*}[n]$ can still be efficiently found. In fact, the solution follows again a (modified) waterfilling scheme $p_{k,2}^{m*}[n]=[\frac{\beta^m\log_2(\exp(1))}{\pi_m+\theta_k a_{k,1}[n]h_{k,1}^m[n]}-\frac{1}{h_{k,2}^m}]_0^{\check{p}_{k,2}^m[n]}$; see, e.g., \cite{Sergiy_icassp11}. Differently, when all four terms in \eqref{E:opt_ind_case1} are considered, the optimization is challenging because $\varphi_k^m(\cdot)$ is not concave any more. The reason is that the last term is strictly convex, rendering the sum of the four terms in \eqref{E:opt_ind_case1} non-concave and therefore, the optimization non-convex.

However, the fact of the optimization not being convex does not necessarily imply that $p_{k,2}^{m*}[n]$ cannot be efficiently found. The first reason is that optimizing $\varphi_k^m(\cdot)$ involves a single (scalar) variable. As a result, simple line search methods can be used. The second reason is that the structure of $\varphi_k^m(\cdot)$ can be exploited to focus the search on a small region. For example, it can be rigorously shown that the waterfilling solution is an upperbound for $p_{k,2}^{m*}[n]$. Moreover, if the CSI is perfect, then $\varphi_k^m(\cdot)$ has at most three stationary points, so that $p_{k,2}^{m*}[n]$ is either 0 or one of those three points. Once $\{p_{k,2}^{m*}[n]\}_{m=1}^M$ are found, finding $\{w_{k,2}^{m*}[n]\}_{m=0}^M$ just requires the evaluation of closed-form expressions [cf. \eqref{E:opt_sched_case1}]. In other words, because in the dual domain the problem can be separated across users and channels, optimizing the Lagrangian does not require solving one non-convex problem in a $(2M+1)K$ dimensional space. Rather, $(M+1)K$ closed forms need to be evaluated (for the scheduling coefficients), and $MK$ non-convex problems in a \emph{one-dimensional space} need to be solved (for the power loadings).

The expressions obtained in this section revealed how the optimal RA depends on the (perfect) CSI and the Lagrange multipliers. Schemes to compute the multipliers in our CR setup are discussed in the next section, while the alternatives to account for CSI imperfections are analyzed in Sec. \ref{S:ImperfectCSI}.

\section{Stochastic estimation of the multipliers}\label{S:LagrangeMultipliers}

Different methods can be used to obtain the value of $\pi^m$, $\theta_k$ and $\rho_k$. Based on Lagrangian Duality Theory, $\{\pi^m,\theta_k,$ $\rho_k\}$ are set to a constant value $\{\pi^{m*},\theta_k^*,\rho_k^*\}$ corresponding to the value that maximizes the dual function associated with \eqref{E:optRA}. Since our problem has zero duality gap, when $\pi^m=\pi^{m*}$, $\theta_k=\theta_k^*$ and $\rho_k=\rho_k^*$ are substituted into \eqref{E:opt_ind_case1}-\eqref{E:opt_sched_case1}, the resulting RA is the optimal solution of \eqref{E:optRA} \cite{Bertsekas_Book_CA03}. To find such values, one has to resort to iterative search algorithms such as dual subgradient methods \cite{Bertsekas_Book_CA03}, which at each iteration update the value of the multiplier according to the long-term violation of the corresponding constraint (let us recall that regardless of the convexity of the primal problem, the dual problem is always convex). Dual subgradient methods (either with constant or diminishing stepsize) and dual descend methods are reasonable alternatives for the problem at hand. Methods exploiting the separability in the dual domain can be used too. The main drawback associated with all previous methods is that at every iteration, the expectations in the long-term constraints (which require averaging over all possible states of $\mathbf{h}$) need to be computed. Moreover, the multipliers have to be recomputed if either the long-term distribution of the channels or the number of users change.

Recently, alternative approaches that rely on stochastic approximation tools have been proposed to find the value of the multipliers \cite{AleTSP11,quantizedCR,Joint_S_RA_Xin10}. These approaches do not try to find the optimal value of $\{\pi^{m*},\theta_k^*,\rho_k^*\}$, but time-varying estimates of them $\{\pi^m[n],\theta_k[n],$ $\rho_k[n]\}$ which are updated at every instant $n$ and remain sufficiently close to $\{\pi^{m*},\theta_k^*,\rho_k^*\}$. An important advantage of these approaches is that their computational complexity is very low. Moreover, they exhibit additional advantages that are especially attractive in CR setups. Namely: i) they are robust to channel non-stationarities (which may arise in environments with interference); ii) they do not need to have statistical knowledge of the channels; and iii) they can cope with changes in either the secondary network (number of users, or QoS levels) or the primary network (limits on the interference power, rate loss, or capacity function of the PUs). In other words, stochastic schemes offer a way to learn the environment online and keep track of its time variation. As we will see, the only price to pay is that the resulting schemes are slightly suboptimal.

To be specific and rigorous, with $\eta_{\pi}$, $\eta_{\theta}$ and $\eta_{\rho}$ being small and constant stepsizes, the following iterations are proposed
\begin{eqnarray}
\label{E:stoch_pow}
\hspace{-.2cm}\pi^m[n+1]\hspace{-.15cm}&=&\hspace{-.15cm}\left[\pi^{m{{}^{}}^{{}^{}}}[n]-\eta_{\pi}(\check{p}^m\hspace{-.07cm}- \hspace{-.13cm}\sum\nolimits_{ k}w_{k,2}^{m*}[n]p_{k,2}^{m*}[n] )\right]_0^{\infty}\\
\nonumber
\hspace{-.2cm}\theta_k[n\hspace{-.01cm}+\hspace{-.1cm}1]\hspace{-.15cm}&=&\hspace{-.15cm}\Big[\theta_k[n]
-\eta_{\theta}a_{k,1}[n]\Big(\check{p}_{k,1}\\
\label{E:stoch_interf_over}
\hspace{-.15cm}&-&\hspace{-.15cm}\hspace{-.1cm} \sum_{m}\hspace{-.01cm}w_{k,2}^{m*}[n]h_{k,1}^m[n]p_{k,2}^{m*}[n]\Big)\Big]_0^{\infty}\\
\nonumber
\hspace{-.2cm}\rho_k[n\hspace{-.1cm}+\hspace{-.1cm}1]\hspace{-.15cm}&=&\hspace{-.15cm}\Big[\rho_k[n]\hspace{-.001cm}+
\eta_{\rho}a_{k,1}[n]\Big(\check{r}_{k,1}\\
\label{E:stoch_interf}\hspace{-.15cm}&-&\hspace{-.15cm} {\sum}_{m}\hspace{-.001cm}w_{k,2}^{m*}[n]r_{k,1}(h_{k,1}^m[n]p_{k,2}^m[n])\Big)\Big]_0^{\infty}.
\end{eqnarray}
From an optimization point of view, the updates in \eqref{E:stoch_pow}-\eqref{E:stoch_interf} form an
unbiased stochastic subgradient of the dual function of \eqref{E:optRA}; see \cite{Bertsekas_Book_CA03}. Assuming that the updates in \eqref{E:stoch_pow}-\eqref{E:stoch_interf} are bounded, the following optimality/feasibility result can be shown\footnote{A proof of this result is not presented here due to space limitations, but it can be derived following the lines of \cite{AleTSP11,am_etal_tvt12}.}.
\begin{proposition}
The sample average of the stochastic RA: i) is feasible and ii) entails a small loss of performance relative to the optimal solution of \eqref{E:optRA}. Specifically, defining $\eta:=\max\{\eta_{\pi}, \eta_{\theta}, \eta_{\rho}\}$;  $\bar{p}_2^m[n]$$:=\frac{1}{n}$$\sum_{l=1}^n\sum_k w_{k,2}^{m*}[l]p_{k,2}^{m*}[l]$;  $\bar{c}_2[n]:=$ $\frac{1}{n}$$\sum_{l=1}^n\sum_{k,m} $ $\beta^m w_{k,2}^{m*}[l]$$r_{k,2}^m$ $(h_{k,2}^m[l],$$p_{k,2}^{m*}[l])$;  $\bar{p}_{k,1}[n]$$:=\frac{1}{n}$ $\sum_{l=1}^na_{k,1}[l]\sum_{m}$ $w_{k,2}^{m*}[l]$$h_{k,1}^m[l]p_{k,2}^{m*}[l]$; and $\bar{r}_{k,1}[n]$$:=\frac{1}{n}$$\sum_{l=1}^na_{k,1}[l]$$\sum_{m}$$w_{k,2}^{m*}[l]$$r_{k,1}(h_{k,1}^m[l]p_{k,2}^{m*}[l])$. Then, it holds with probability one that as $n\rightarrow \infty$:\\
$\hspace{.5cm}$i) $\bar{p}_2^m[n]\leq\check{p}^m$, $\bar{p}_{k,1}[n]\leq\check{p}_{k,1}$, $\bar{r}_{k,1}[n]\geq\check{r}_{k,1}$, and \\
$\hspace{.5cm}$ii) $\bar{c}_2[n]\geq \bar{c}_2^*-\Delta(\eta)$, where $\Delta(\eta)\rightarrow 0$ as $\eta\rightarrow 0$.
\end{proposition}
In words, the proposition guarantees asymptotic optimality of the stochastic iterates because they give rise to a RA which is feasible and achieves a value (performance) arbitrarily close to $\bar{c}_2^*$, which is the optimal objective that the original (non-stochastic) solution of \eqref{E:optRA} achieves [cf. \eqref{E:optRA_obj}]. Note also that $\eta$ can be used as a parameter to set the tradeoff between optimality and tracking capabilities. If optimality is the only concern, the stochastic iterations in \eqref{E:stoch_pow}-\eqref{E:stoch_interf} could be run using a time-varying stepsize $\eta[n]$ which diminishes with time. Under mild conditions, it can be shown that such iterations converge to the optimal solution; see, e.g., \cite{quantizedCR} for details. Clearly, the price to pay in that case is that the algorithms would lose their tracking capabilities.

\begin{remark}
In this work, we have assumed that there is a central scheduler (AP) that gathers the CSI, finds the optimum RA, and runs the stochastic iterates. Moreover, we have also assumed that the signalling channels which convey the control information are error free. Nonetheless, it is worth remarking that the stochastic estimates are robust to errors. In fact, if the errors in the updates are bounded and have zero mean, then the results in Prop. 1 still hold. See \cite{GatsisErrorsExchanges} for a related result. In addition, the next section will show that our schemes are also robust to errors/imperfections in the CSI.
\end{remark}

\section{Imperfect channel state information}\label{S:ImperfectCSI}

The optimal RA schemes were designed assuming that the CSI was perfect. Here, we relax that assumption and account for CSI imperfections. Although the assumption of perfect CSI may be reasonable for some wireless systems, it is unlikely to hold in CR scenarios (see related discussion in Sec. \ref{S:Introduction}). This is especially true for the CSI of the primary network, which is typically more difficult to obtain and entails a higher cost than that of secondary links. We first present different alternatives to model the CSI imperfections and then, describe how the RA schemes have to be modified to account for them.

The main change in the formulation when the CSI is not perfect is that the values of $a_{k,1}[n]$, $h_{k,1}^m[n]$ and $h_{k,2}^m[n]$ (instantaneous CSI) are not longer deterministically known at instant $n$. Rather, the knowledge of $a_{k,1}[n]$, $h_{k,1}^m[n]$ and $h_{k,2}^m[n]$ will be probabilistic and time varying. As a result, the CSI now will correspond to the probability density function (pdf) of $a_{k,1}[n]$, $h_{k,1}^m[n]$, $h_{k,2}^m[n]$ available at time $n$. Such a pdf will be referred to as instantaneous \emph{belief} and denoted as $b_{k,1}(x~|n)$, $b_{k,1}^m(x~|n)$, $b_{k,2}^m(x~|n)$, respectively. The specific expression for the instantaneous belief will depend on the operating conditions of the system. Focusing on $h_{k,1}^m[n]$ for illustrative purposes, two extreme examples are analyzed next. First, consider the case when the CSI is perfect. For this case, the value of $h_{k,1}^m[n]$ at instant $n$ is perfectly known, so that belief at instant $n$ (instantaneous pdf) would be $b_{k,1}^m(x~|n)=\delta(x-h_{k,1}^m[n])$, where $\delta(\cdot)$ is a Dirac delta function. Consider now that no instantaneous measurements are available, so that only (long-term) statistical CSI is available. For the case of Rayleigh channels, the belief would be $b_{k,1}^m(x~|n)=\exp(x/\bar{h}_{k,1}^m)/\bar{h}_{k,1}^m$, where $\bar{h}_{k,1}^m$ represents the average gain of the SU-to-PU channel. Clearly, in this case the belief would not vary with time.

Three different sources of imperfections are considered here: quantized CSI, noisy CSI, and outdated CSI. For each of them, we first give a high level description of how to model the imperfections and the corresponding belief. Then, we provide several examples that will allow us to gain insights and be more specific. Regarding the first source of imperfections, research has consistently shown that feedbacking a small number of information bits about the instantaneous channel conditions to the transmitter (or schedulers) can allow near optimal channel adaptation \cite{SI_jsac_lrf}. To implement such schemes, the channel domain has to be quantized into non-overlapping quantization regions. Such quantization can be carried out jointly for different channels (vector quantization) or separately for each of them. Once the quantizer is known, at each instant the transmitter is notified of the region the instantaneous channels falls into. The instantaneous belief will be given by the pdf of the channel gain within the active region. A different source of imperfections is the presence of noise in the channel measurements. A zero-mean additive white noise is typically assumed for the noise, so that the belief will be given by the instantaneous channel measurement and the noise pdf. Many systems do not estimate the power gain of the channel, but its complex low-pass equivalent. In such a case, the (complex) noise would affect the low-pass equivalent. The belief in this case can be obtained from the actual measurement, the noise distribution and taking into account that power gain is the squared modulus of the complex low-pass equivalent. Finally, we also consider that the CSI may be outdated. This model is well motivated in CRs where sensing the (PU) channels entails a high cost so that they are cannot be sensed at every time instant. To update the belief in this case we need to assume a specific time-correlation model for the CSI. Based on that model and on the available measurements up to instant $n$, the belief is estimated using stochastic prediction/correction schemes.

\begin{example}
A simple but very effective alternative to define the quantized CSI is to use a scalar quantizer for each of the channel gains. For example, focusing on the SU-to-SU channels, the domain of $h_{k,2}^m[n]$ can be divided into $L$ non overlapping intervals $[\tau_{k,2}^{m,l-1},\tau_{k,2}^{m,l})$, where $l=0,\ldots,L$, $\tau_{k,2}^{m,l}$ stands for the $l$th quantization threshold and $\tau_{k,2}^{m,0}=0$ and $\tau_{k,2}^{m,L}=\infty$. Clearly, in this case $\log_2(L)$ bits suffice to identify the region (interval) channel $h_{k,2}^m[n]$ falls into. Most quantized CSI designs ignore the time-correlation of the channel and assume that the CSI is available instantaneously and free of errors  \cite{SI_jsac_lrf}. In such a scenario, let $l_{k,2}^m[n]$ be the index which identifies the region the channel $h_{k,2}^m[n]$ falls into. If the channel $h_{k,2}^m[n]$ follows a exponential distribution (Rayleigh model) and its average gain is $\bar{h}_{k,2}^m$, then the belief of $h_{k,2}^m[n]$ at instant $n$ is $b_{k,2}^m(x~|n)=[\exp(-x/\bar{h}_{k,2}^m)/\bar{h}_{k,2}^m]/\Pr\{h_{k,2}^m \in [\tau_{k,2}^{m,l-1},\tau_{k,2}^{m,l})\}$.
\end{example}

\begin{example}
The task of acquiring the Boolean variable $a_{k,1}[n]$ is basically a detection problem. Consider that the output of the detection process is binary and denoted by $\tilde{a}_{k,1}[n]$. In order to incorporate the sensing errors into our model, we denote the probabilities of miss detection and false alarm as $P_{MD}:=\Pr\{\tilde{a}_{k,1}[n]\hspace{-.05cm}=\hspace{-.05cm}0\hspace{.1cm}|a_{k,1}[n]\hspace{-.05cm}=\hspace{-.05cm}1\}$ and $P_{FA}:=\Pr\{\tilde{a}_{k,1}[n]\hspace{-.05cm}=\hspace{-.05cm}1\hspace{.1cm}|a_{k,1}[n]\hspace{-.05cm}=\hspace{-.05cm}0\}$, respectively. Based on those, we define $P_{0|0}:=[(1-P_{FA})P_0]/[(1-P_{FA})P_0+P_{MD}P_1]$ and $P_{1|1}:=[(1-P_{MD})P_1]/[P_{FA}P_0+(1-P_{MD})P_1]$, where $P_0$ and $P_1$ stand for the long-term probabilities of $\Pr\{\tilde{a}_{k,1}\hspace{-.05cm}=\hspace{-.05cm}0\}$ and $\Pr\{\tilde{a}_{k,1}\hspace{-.05cm}=\hspace{-.05cm}1\}$, respectively. If the time-correlation of $a_{k,1}[n]$ is ignored, then the belief of $a_{k,1}[n]$ at time $n$ is simply: $b_{k,1}(x~|n)=P_{0|0}\delta(x)+(1-P_{0|0})\delta(x-1)$ if $\tilde{a}_{k,1}[n]=0$; and $b_{k,1}(x~|n):=(1-P_{1|1})\delta(x)+P_{1|1}\delta(x-1)$ if $\tilde{a}_{k,1}[n]=1$. Schemes to update the belief for more general sensing models and that leverage the time-correlation of the PUs activity can be found in, e.g., \cite{JSAC}.
\end{example}

\begin{example}
In this example, we design prediction/correction schemes for a practical channel/measurement model for the SU-to-PU channels. Let $g_{k,1}^m[n]$ be the low-pass equivalent of the SU-to-PU channel, so that $h_{k,1}^m[n]=|g_{k,1}^m[n]|^2$. We will assume that $g_{k,1}^m[n]$ is a complex Gaussian process with independent real and imaginary parts (Rayleigh model). For notational convenience we will deal with $g_{k,1}^m[n]$ as a $2\times 1$ vector whose first and second entries correspond to the real and imaginary parts, respectively. The time dynamics of $g_{k,1}^m[n]$ are assumed to follow a first-order Markovian model with $g_{k,1}^m[n]=(\varrho_k^m)^{1/2}g_{k,1}^m[n-1]+(1-\varrho_k^m)^{1/2}d_{k,1}^m[n]$ where $\varrho_k^m$ represents the autocorrelation coefficient and $d_{k,1}^m[n]$ an innovation process independent of $g_{k,1}^m[n]$. The process $d_{k,1}^m[n]$ is assumed to be white and complex Gaussian distributed with zero mean and diagonal covariance matrix $\frac{1}{2} \mathbf{I}_2$, where $\mathbf{I}_2$ is the $2\times 2$ identity matrix \cite{GoldsmithBook}.
Once the model of the ground-truth channel has been described, we introduce the model for the measurements and errors. For such a purpose, let $s_k^m[n]$ denote a Boolean variable which is one if the channel $g_{k,1}^m$ is sensed at instant $n$ and zero otherwise. Moreover, let $\tilde{g}_{k,1}^m[n]$ denote the noisy measurement of $g_{k,1}^m[n]$ obtained if $s_k^m[n]=1$. The measurement is modeled as $\tilde{g}_{k,1}^m[n]=g_{k,1}^m[n]+v_k^m[n]$ where $v_k^m[n]$ is a white noise independent of $g_{k,1}^m[n]$ which  follows a complex Gaussian distribution with zero mean and diagonal covariance matrix $\nu_k^m\mathbf{I}_2$. Let $f_{g_{k,1}^m[n]}(x)$ denote the pdf of $g_{k,1}^m[n]$ at instant $n$, conditioned to all measurements up to instant $n$. Under the previous model, it readily follows that $f_{g_{k,1}^m[n]}(x)$ is Gaussian pdf and its mean and covariance (denoted, respectively, as $\mu_k^m[n]$ and $\upsilon_k^m[n]$) suffice to describe the full distribution. The stochastic iterations to update $\mu_k^m[n]$ and $\upsilon_k^m[n]$ are described next.

If $s_k^m[n]=0$, then it holds that $\mu_k^m[n]=(\varrho_k^m)^{1/2}\mu_k^m[n-1]$ and $\upsilon_k^m[n]=\varrho_k^m\upsilon_k^m[n-1]+(1-\varrho_k^m)\frac{1}{2}\mathbf{I}_2$. If $s_k^m[n]=1$, we first update the belief of the previous instant to get the predictions $\hat{\mu}_k^m[n]=(\varrho_k^m)^{1/2}\mu_k^m[n-1]$ and $\hat{\upsilon}_k^m[n]=\varrho_k^m\upsilon_k^m[n-1]+(1-\varrho_k^m)\frac{1}{2}\mathbf{I}_2$. Then, we use the measurement $\tilde{g}_k^m[n]$ to correct the predictions as follows:
\vspace{-.02cm}
\begin{eqnarray}
\hspace{-.1cm}\mu_k^m[n]\hspace{-.1cm}&=&\hspace{-.1cm}(\hat{\upsilon}_k^m[n]+\nu_k^m)^{-1}(\hat{\upsilon}_k^m[n] \tilde{g}_k^m[n] + \nu_k^m \hat{\mu}_k^m[n])\\
\hspace{-.1cm}\upsilon_k^m[n]\hspace{-.1cm}&=& \hspace{-.1cm}(\hat{\upsilon}_k^m[n]+\nu_k^m)^{-1}(\hat{\upsilon}_k^m[n]\nu_k^m).
\end{eqnarray}

Clearly, when $s_k^m[n]=1$ the updates correspond to those of a classical Kalman filter. Different prediction/correction steps will be required if either the time dynamics or the sensing errors are modeled differently. See, e.g., \cite{JSAC} for alternative models. As mentioned before, based on $f_{g_{k,1}^m[n]}(x)$ (instantaneous pdf of $g_{k,1}^m[n]$), the belief $b_{k,1}^m(x~|n)$ (instantaneous pdf of $h_{k,1}^m[n]$) can be obtained by using the transformation $h_{k,1}^m[n]=|g_{k,1}^m[n]|^2$.
\end{example}

To finish this section, we introduce notation $\mathbf{\tilde{h}}[n]$ to denote the overall imperfect CSI available at time $n$. For example, suppose that: a) the CSI of the SU-to-SU gains is quantized as described in Example 1; b) the errors on the activity of the PUs follow the model described in Example 2; and c) the CSI of the SU-to-PU channels is outdated and noisy as described in Example 3. With these operating conditions, $\mathbf{\tilde{h}}[n]$ is a vector of length $(3M+1)K$ containing: a) the region index of each of the gains of the $MK$ SU-to-SU links; ii) the probability of each of the $K$ PUs being active; and iii) the means and variances of the $MK$ SU-to-PU links. Clearly, based on the information gathered on $\mathbf{\tilde{h}}[n]$, the instantaneous beliefs $b_{k,1}(x~|n)$, $b_{k,1}^m(x~|n)$, $b_{k,2}^m(x~|n)$ can be trivially obtained. For notational convenience, we will use $\mathbf{b}(x~|n)$ to denote the belief of the CSI of the \emph{overall} system. Moreover, $\mathbf{b}(x~|n)$ will be written as $\mathbf{b}(x~|\mathbf{\tilde{h}}[n])$ whenever is convenient to stress the dependence on $\mathbf{\tilde{h}}[n]$.

\subsection{Modifying the RA schemes}
The first step to design RA schemes capable of accounting for CSI imperfections is to modify the formulation of the constraints which depend explicitly on the instantaneous CSI. Strictly speaking, the formulation of the long-term constraints in \eqref{E:optRA_c_av_pow}, \eqref{E:optRA_c_av_power_interf} and \eqref{E:optRA_c_av_rate_loss} (and the objective $\bar{c}_2$) do not have to be modified. One just has to take into account that the total expectation $\mathbbm{E}_{\mathbf{h}}[\cdot]$ in those constraints can be rewritten as $\mathbbm{E}_{\mathbf{\tilde{h}}}[\mathbbm{E}_{\mathbf{b}(x|\mathbf{\tilde{h}})}[\cdot]]$. The notation emphasizes that the inner expectation is taken over $a_{k,1}[n]$, $h_{k,1}^m[n]$ and $h_{k,2}^m[n]$ according to the pdfs in $\mathbf{b}(x|\mathbf{\tilde{h}})$.
Differently, the short-term constraints in \eqref{E:optRA_c_inst_power_interf_ind_users} and \eqref{E:optRA_c_inst_rate_loss_ind_users} need to be modified. When the CSI is imperfect, those constraints involve random variables, so that strict satisfaction of the constraints may be impossible (e.g., if the instantaneous belief has infinite support). As a result, the constraints have to be reformulated. A reasonable reformulation is to take \emph{expectations across the instantaneous belief} at both sides of the constraints and consider
\begin{eqnarray}
\nonumber
\mathbbm{E}_{b_{k,1}(x|n)}[a_{k,1}[n]]\mathbbm{E}_{b_{k,1}^m(x|n)}\left[h_{k,1}^m[n] \right]p_{k,2}^m[n]\\
\label{E:optRA_c_inst_power_interf_ind_users_imp}\leq \mathbbm{E}_{b_{k,1}(x|n)}[a_{k,1}[n]]\check{p}_{k,1},\\
\nonumber\mathbbm{E}_{b_{k,1}(x|n)}[a_{k,1}[n]] \mathbbm{E}_{b_{k,1}^m(x|n)}\left[r_{k,1}(h_{k,1}^m[n] p_{k,2}^m[n])\right]\\
\label{E:optRA_c_inst_rate_loss_ind_users_imp}\geq \mathbbm{E}_{b_{k,1}(x|n)}[a_{k,1}[n]]\check{r}_{k,1}.
\end{eqnarray}
Note that to gain intuition in \eqref{E:optRA_c_inst_power_interf_ind_users_imp} and \eqref{E:optRA_c_inst_rate_loss_ind_users_imp} we have implicitly assumed that $a_{k,1}[n]$ and $h_{k,1}^m[n]$ are independent, so that the expectations were obtained separately. The long-term expectations in \eqref{E:optRA_c_av_power_interf} and \eqref{E:optRA_c_av_rate_loss} are different from those in \eqref{E:optRA_c_inst_power_interf_ind_users_imp} and \eqref{E:optRA_c_inst_rate_loss_ind_users_imp}. In the former, the expectations were taken considering all time instants. In the latter, the expectations are taken at instant $n$ and only over the CSI uncertainties. Clearly, as the knowledge of the CSI improves, the beliefs approximate to a Dirac delta centered in the actual value of the channel and hence, the constraints in \eqref{E:optRA_c_inst_power_interf_ind_users_imp} and \eqref{E:optRA_c_inst_rate_loss_ind_users_imp} approximate to those in \eqref{E:optRA_c_inst_power_interf_ind_users} and \eqref{E:optRA_c_inst_rate_loss_ind_users}. As we did in Sec. \ref{Ss:CRConstraints}, to handle the short-term DSA constraints we solve \eqref{E:optRA_c_inst_power_interf_ind_users_imp} and \eqref{E:optRA_c_inst_rate_loss_ind_users_imp} w.r.t. $p_{k,2}^m[n]$ and redefine the maximum instantaneous peak power constraint as $\check{\tilde{p}}_{k,2}^m[n]:=\min\{\tilde{x}_k^m[n],\tilde{y}_k^m[n],\check{p}_{k,\max}^m\}$, where $\tilde{x}_k^m[n]$ and $\tilde{y}_k^m[n]$ are the roots of \eqref{E:optRA_c_inst_power_interf_ind_users_imp} and \eqref{E:optRA_c_inst_rate_loss_ind_users_imp}, respectively. Another reasonable reformulation to handle the CSI imperfections is to consider that \eqref{E:optRA_c_inst_power_interf_ind_users} and \eqref{E:optRA_c_inst_rate_loss_ind_users} need to hold with a certain short-term probability (e.g., the probability of the interference power at time $n$ exceeding $\check{p}_{k,1}$ has to be less than a certain value). The procedure to deal with the constraints would be similar. The instantaneous belief would be used to solve the constraints w.r.t. the $p_{k,2}^m[n]$, the corresponding values of $\tilde{x}_k^m[n]$ and $\tilde{y}_k^m[n]$ would be found, and such values would be used to obtain $\check{\tilde{p}}_{k,2}^m[n]$.

With these modifications in mind, it can be shown (see App. A) that the optimal RA with imperfect CSI is
\begin{eqnarray}
\label{E:opt_ind_case1_imp}\tilde{\varphi}_k^m(p_{k,2}^m[n])&:=&\mathbbm{E}_{\mathbf{b}(x|n)}\left[\varphi_k^m(p_{k,2}^m[n])\right],\hspace{.3cm}\\
\label{E:opt_pow_case1_imp}p_{k,2}^{m*}[n]&:=&\left[\arg\max_{p_{k,2}^m[n]}~\tilde{\varphi}_k^m(p_{k,2}^m[n])\right]_0^{\check{\tilde{p}}_{k,2}^m[n]}\\
\nonumber w_{k,2}^{m*}[n]&:=&\mathbbm{1}_{\{ m=\arg\max_{l} \tilde{\varphi}_k^l(p_k^{l*}[n])\}}\\
\label{E:opt_sched_case1_imp}&~&\cdot\mathbbm{1}_{\{p_{k,2}^{m*}[n]>0~\vee~m=0\}}.\hspace{.3cm}
\end{eqnarray}
In most practical scenarios, the SU-to-SU channels are statistically independent of the SU-to-PU channels. The same holds true for the activity of the PUs. In such a case, the indicator in \eqref{E:opt_ind_case1_imp} can be written as $\tilde{\varphi}_k^m(p_{k,2}^m[n])=\beta^m \mathbbm{E}_{b_{k,2}^m(x|n)}[r_{k,2}^m(h_{k,2}^m[n]p_{k,2}^m[n])]-\pi^mp_{k,2}^m[n]-\theta_k \mathbbm{E}_{b_{k,1}(x|n)}\left[a_{k,1}[n]\right]\mathbbm{E}_{b_{k,1}^m(x|n)}[h_{k,1}^m[n]]p_{k,2}^m[n]+\rho_k \mathbbm{E}_{b_{k,1}(x|n)}\left[a_{k,1}[n]\right]\mathbbm{E}_{b_{k,1}^m(x|n)}[r_{k,1}(h_{k,1}^m[n] p_{k,2}^m[n])]$.
This way, we observe that the fact of having imperfect CSI does not modify the favorable (separable) structure of the optimal RA. The main change is that the optimization in \eqref{E:opt_pow_case1_imp} has to be carried out taking into account the CSI imperfections. In most cases, this will entail a higher computational cost (because the expectations cannot be found in closed form and have to be estimated numerically). If computational burden is a major problem, robust designs that guarantee a worst-case performance and do not require computing expectations are a reasonable alternative.

The last step to account for the CSI imperfections is to modify the schemes that compute the multipliers. If the stochastic schemes in \eqref{E:stoch_pow}-\eqref{E:stoch_interf} are used, a simple way to accomplish that task is to replace the instantaneous updates in the right hand side of \eqref{E:stoch_pow}-\eqref{E:stoch_interf} with their expectations over the instantaneous belief $\mathbf{b}(x|n)$. In such a case, the results in Prop. 1 still hold. In fact, if the expectations over the instantaneous belief were replaced with simple unbiased and bounded estimates, then the results in Prop. 1 would hold too.

\section{Numerical simulations}
The performance of our schemes is analyzed here via numerical simulations. Since the schemes are optimal, the main purpose is to get insights into the optimal policies and the role of each of the DSA constraints considered. Two test cases are presented. The first one focuses on the overall sum-capacity performance (optimality) and feasibility of the developed schemes. The effects associated with modification of the interference levels and DSA constraints are analyzed, and perfect CSI is assumed. The second test case analyzes the impact of CSI imperfections.

To simulate challenging propagation conditions for the SUs, the amplitudes of the secondary links are Rayleigh distributed (so that $h_{k,2}^m[n]$ follows an exponential distribution), the average SNR for all users and bands is $3$dB, and the frequency selectivity is assumed to be high, so that gains across bands (sets of subcarriers) are independent. The model for the SU-to-PU links is Rayleigh too, with average gain equal to $0$dB. The gain of the PU-to-PU link is $10$dB and every PU is assumed to be active during 80$\%$ of the time. The remaining parameters are set as follows: $M=5$, $K=10$, $\beta^m=1$, $\check{p}^m=1$, $\check{p}_{k,1}=0.15$, and $\check{\varepsilon}_{k,1}=5\%$. The number of time instants simulated is 20000, the results presented correspond to one single realization of the CSI processes and time averages are calculated discarding the first half of the simulated instants.

\noindent\textbf{Test Case 1: optimality and feasibility.}
To label the schemes in this section, ``A'' stands for average, ``I'' for instantaneous, ``P'' for power and ''C'' for capacity. Seven RA schemes are tested: S1) the optimal scheme that maximizes the performance of the SUs and ignores all DSA constraints (labeled as ``None''); S2) the optimal scheme in this paper considering the long-term interference power constraint \eqref{E:optRA_c_av_power_interf} and the long-term rate loss constraint \eqref{E:optRA_c_av_rate_loss} (labeled as ``APC''); S3) a scheme like APC, but setting $\check{p}_{k,1}=\infty$, i.e., ignoring \eqref{E:optRA_c_av_power_interf} (``AC''); S4) a scheme like APC, but setting $\check{\varepsilon}_k=1$, i.e., ignoring \eqref{E:optRA_c_av_rate_loss} and yielding a scheme very similar to the one in \cite{Sergiy_icassp11} (``AP''); S5) a scheme like AP, but replacing \eqref{E:optRA_c_av_power_interf} with its instantaneous counterpart in \eqref{E:optRA_c_inst_power_interf_ind_users} (``IP''); S6) a scheme like AC, but replacing \eqref{E:optRA_c_av_rate_loss} with its instantaneous counterpart in \eqref{E:optRA_c_inst_rate_loss_ind_users} (``IC''); and S7) a scheme like APC, but replacing both \eqref{E:optRA_c_av_power_interf} and \eqref{E:optRA_c_av_rate_loss} with their instantaneous counterparts \eqref{E:optRA_c_inst_power_interf_ind_users} and \eqref{E:optRA_c_inst_rate_loss_ind_users} (``IPC''). In all cases the CSI is assumed to be error free.

The numerical results corresponding to this test case are plotted in Figs. 1-3. The vertical axes in each of the figures represent the following: in Fig. 1, the long-term weighted sum-capacity of the SUs (denoted as $\bar{c}_2$); in Fig. 2, the long-term interference power at the PUs (the value corresponds to the average across PUs and is denoted as $\bar{p}_1$); and in Fig. 3, the loss on the long-term capacity at the PUs (the value corresponds to the average across PUs and is denoted as $\bar{\varepsilon}_1$). Each of the figures comprises 4 subplots, the horizontal axis in each of the subplots corresponds to the variation of a different parameter: $\bar{h}_{k,1}^m$ (subplot a); $\gamma_k$ (subplot b); $\check{p}_{k,1}$ (subplot c); and $\check{\varepsilon}_{k}$ (subplot d). The long-term power transmitted by the SUs is not plotted because it is always $1$, which is the value set for $\check{p}_2^m$.

\begin{figure}[h]
\vspace{-.4cm}
\centering
\subfloat[Subfigure 1 list of figures text][SNR for SU-to-PU: $\bar{h}_{k,1}^m$]{
\includegraphics[width=0.33\textwidth]{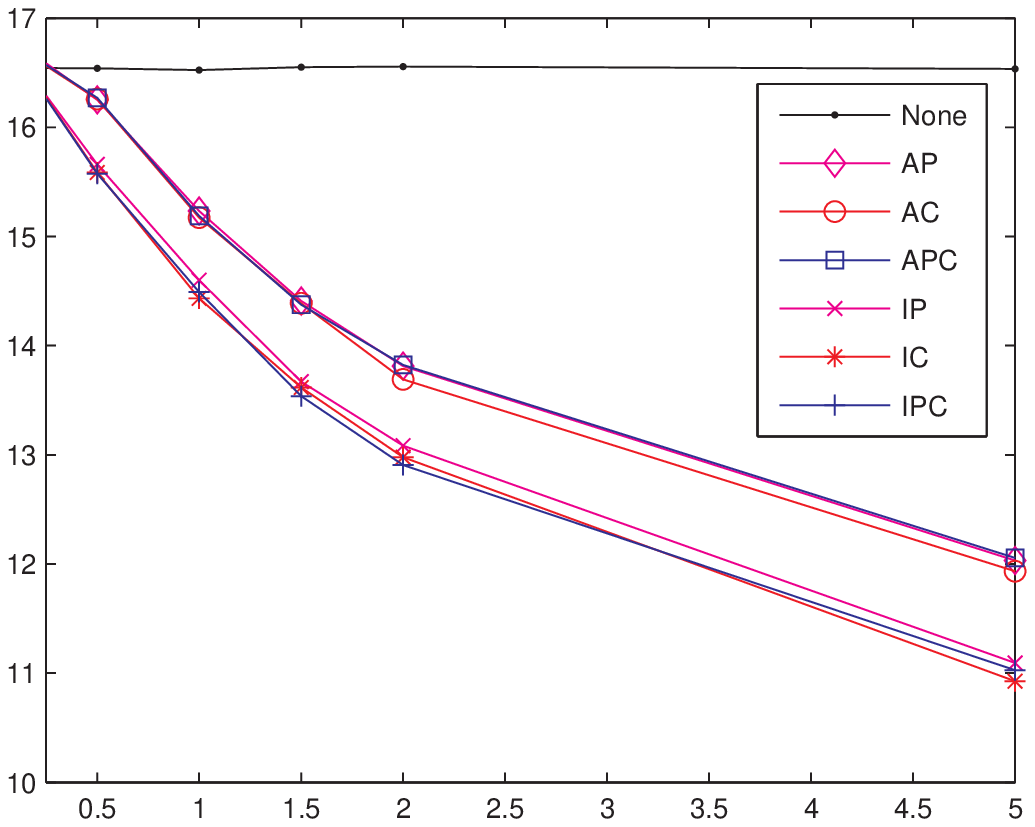}
\label{FigA:subfig1}}
\subfloat[Subfigure 2 list of figures text][SNR for the PU-to-PU: $\gamma_k$]{
\includegraphics[width=0.33\textwidth]{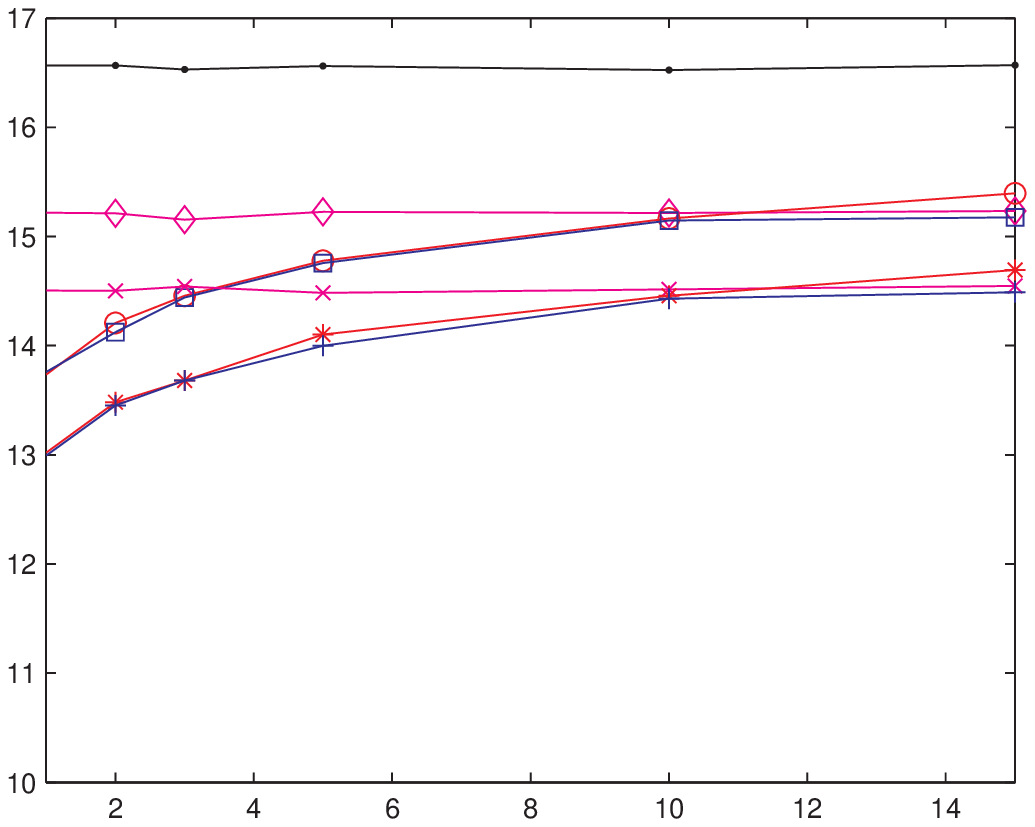}
\label{FigA:subfig2}}\\
\subfloat[Subfigure 3 list of figures text][Max. power at the PUs $\check{p}_{k,1}$]{
\includegraphics[width=0.33\textwidth]{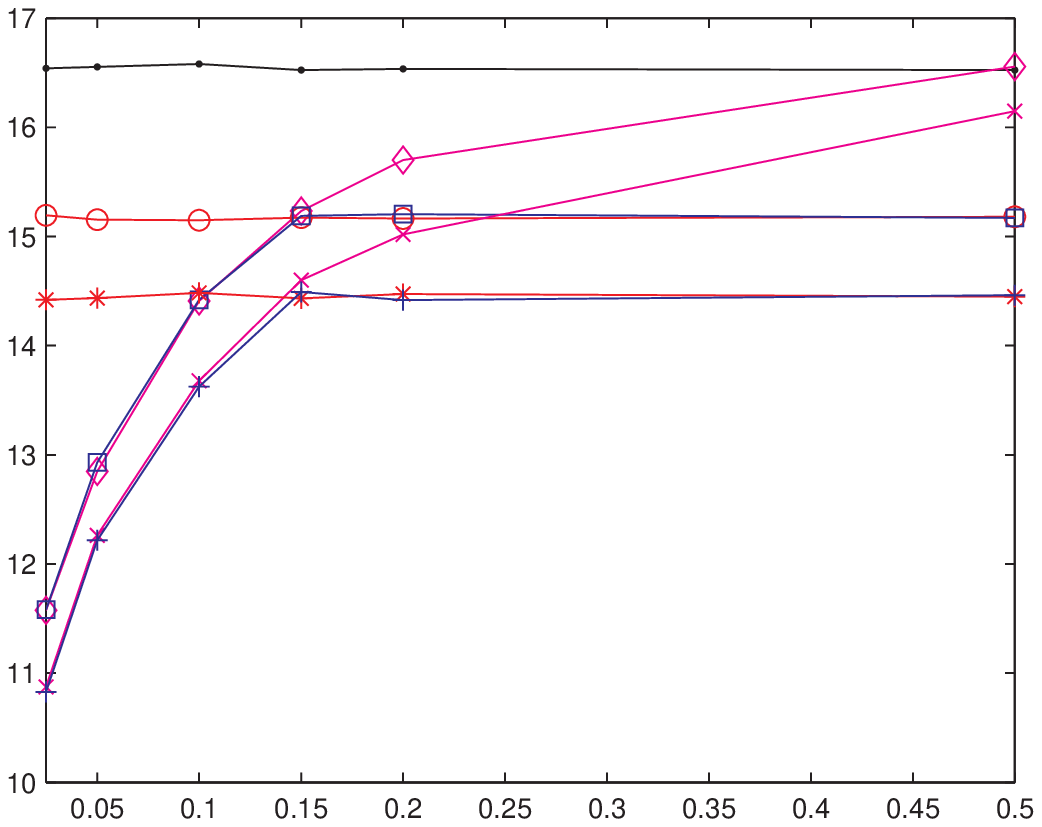}
\label{FigA:subfig3}}
\subfloat[Subfigure 3 list of figures text][Max. capacity loss at PUs: $\bar{\varepsilon}_1$]{
\includegraphics[width=0.33\textwidth]{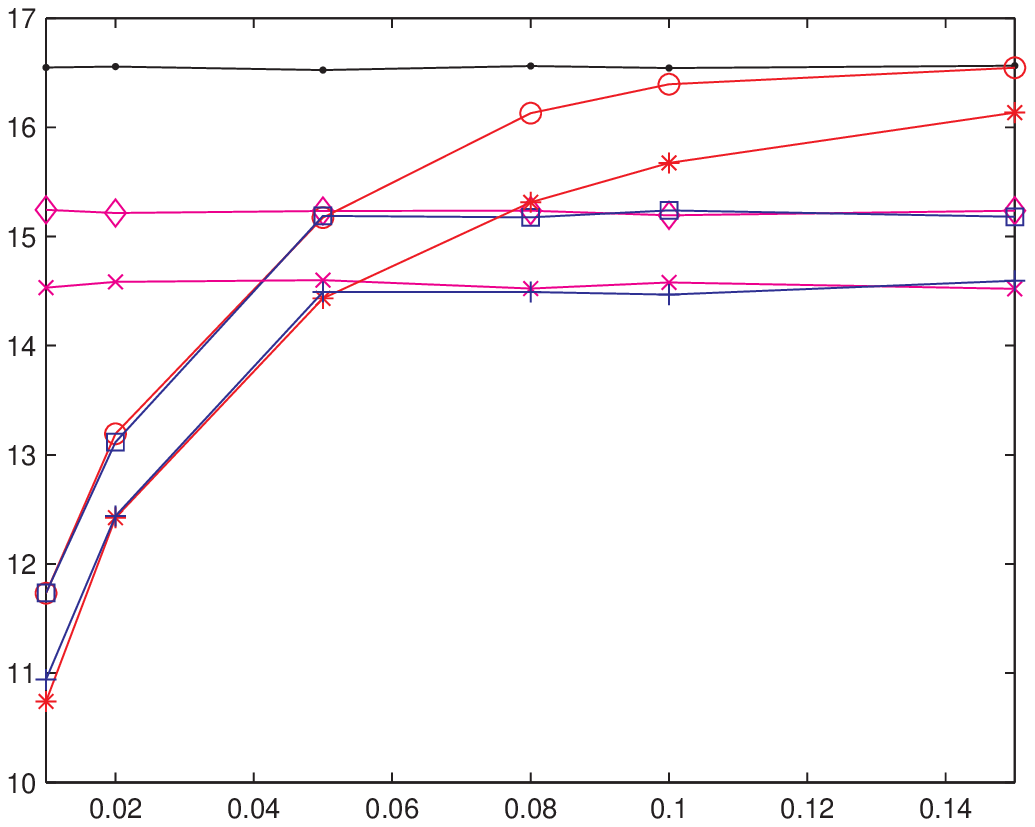}
\label{FigA:subfig4}}
\caption{Variation of $\bar{c}_2$ w.r.t. $\bar{h}_{k,1}^m$, $\gamma_k$, $\check{p}_1$, and $\check{\varepsilon}_1$.}
\label{FigA:globfig}
\end{figure}

\begin{figure}[h]
\vspace{-.4cm}
\centering
\subfloat[Subfigure 1 list of figures text][SNR for SU-to-PU: $\bar{h}_{k,1}^m$]{
\includegraphics[width=0.33\textwidth]{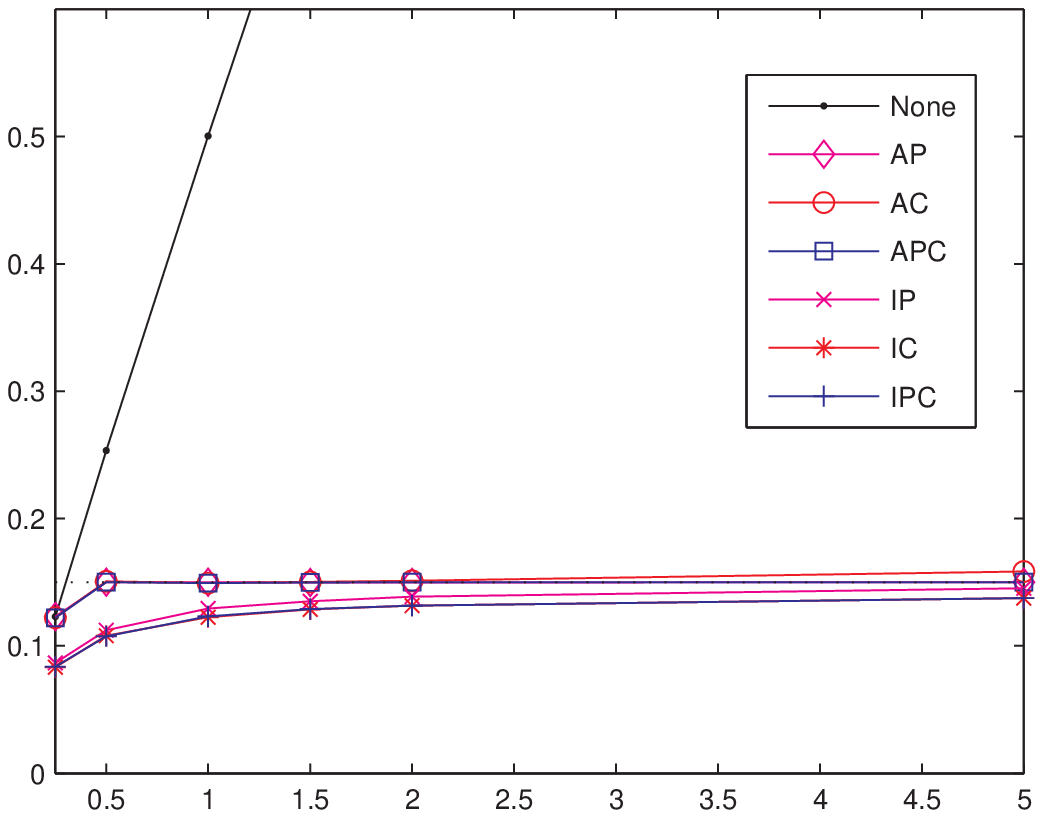}
\label{FigA:subfig1}}
\subfloat[Subfigure 2 list of figures text][SNR for PU-to-PU: $\gamma_k$]{
\includegraphics[width=0.33\textwidth]{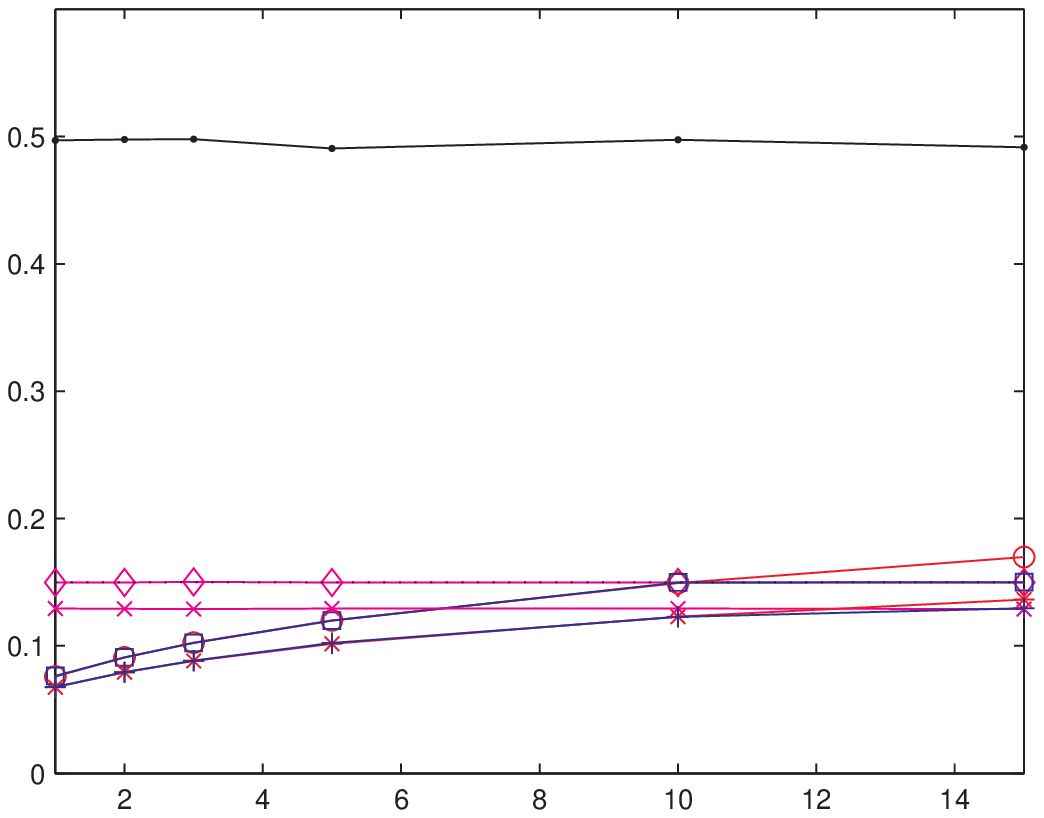}
\label{FigA:subfig2}}\\
\subfloat[Subfigure 3 list of figures text][Max. power at the PUs $\check{p}_{k,1}$]{
\includegraphics[width=0.33\textwidth]{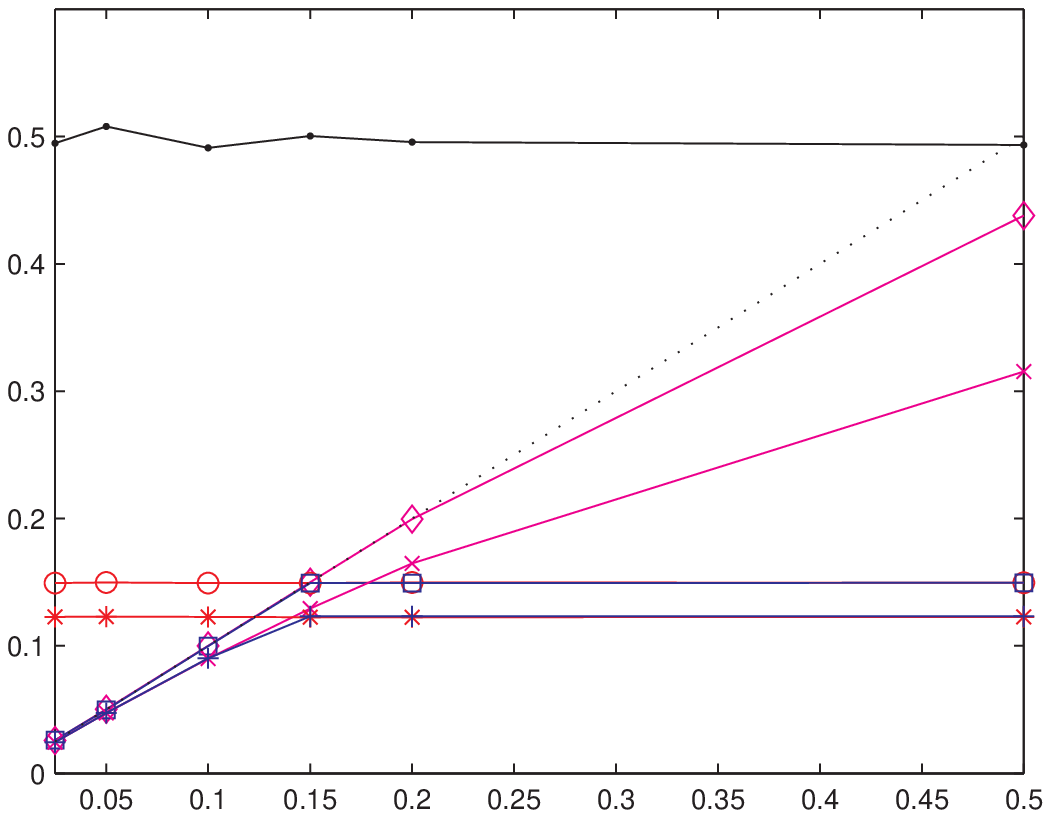}
\label{FigA:subfig3}}
\subfloat[Subfigure 3 list of figures text][Max. capacity loss at PUs: $\bar{\varepsilon}_1$]{
\includegraphics[width=0.33\textwidth]{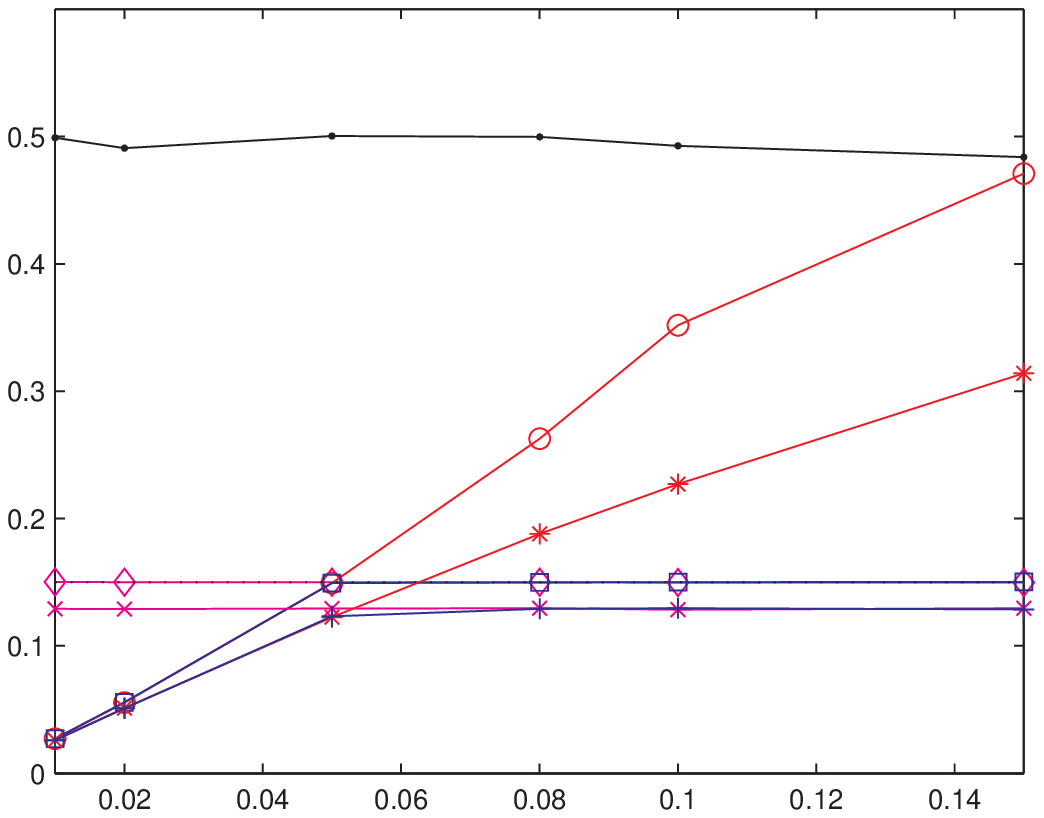}
\label{FigA:subfig4}}
\caption{Variation of $\bar{p}_1$ w.r.t. $\bar{h}_{k,1}^m$, $\gamma_k$, $\check{p}_1$, and $\check{\varepsilon}_1$.}
\label{FigB:globfig}
\vspace{-.6cm}
\end{figure}

\begin{figure}[h]
\vspace{-.4cm}
\centering
\subfloat[Subfigure 1 list of figures text][SNR for SU-to-PU: $\bar{h}_{k,1}^m$]{
\includegraphics[width=0.33\textwidth]{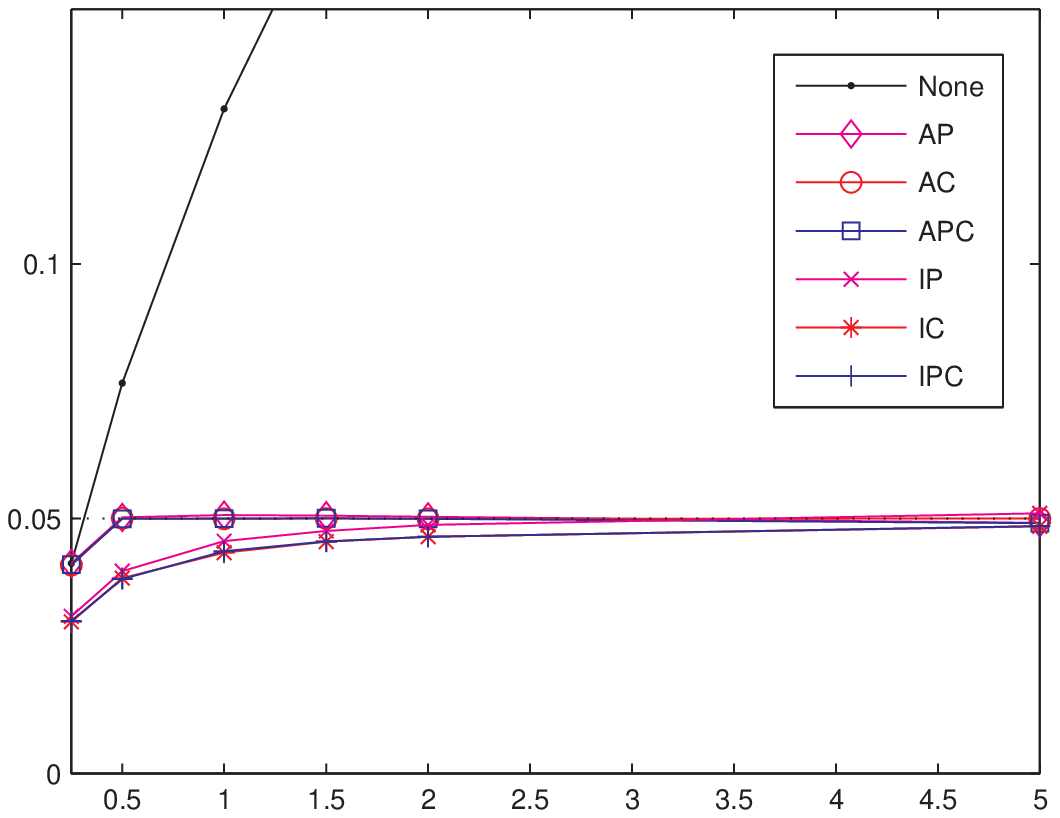}
\label{FigA:subfig1}}
\subfloat[Subfigure 2 list of figures text][SNR for PU-to-PU: $\gamma_k$]{
\includegraphics[width=0.33\textwidth]{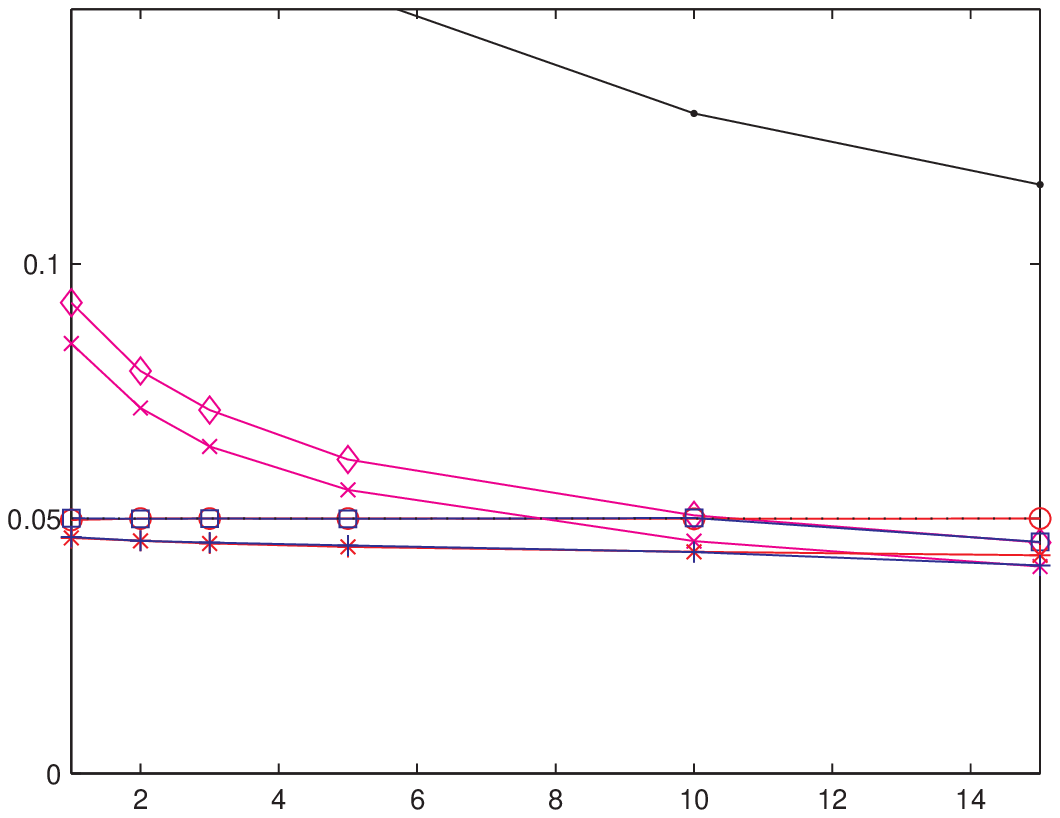}
\label{FigA:subfig2}}\\
\subfloat[Subfigure 3 list of figures text][Max. power at the PUs $\check{p}_{k,1}$]{
\includegraphics[width=0.33\textwidth]{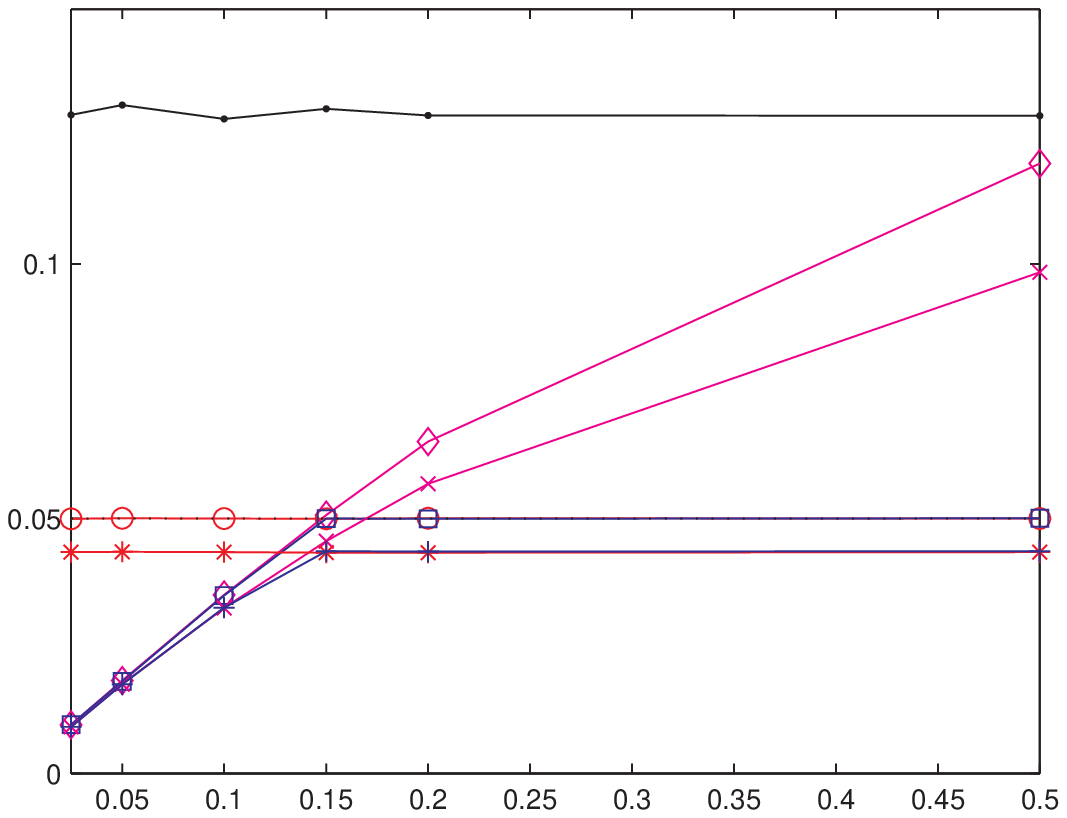}
\label{FigA:subfig3}}
\subfloat[Subfigure 3 list of figures text][Max. capacity loss at PUs: $\bar{\varepsilon}_1$]{
\includegraphics[width=0.33\textwidth]{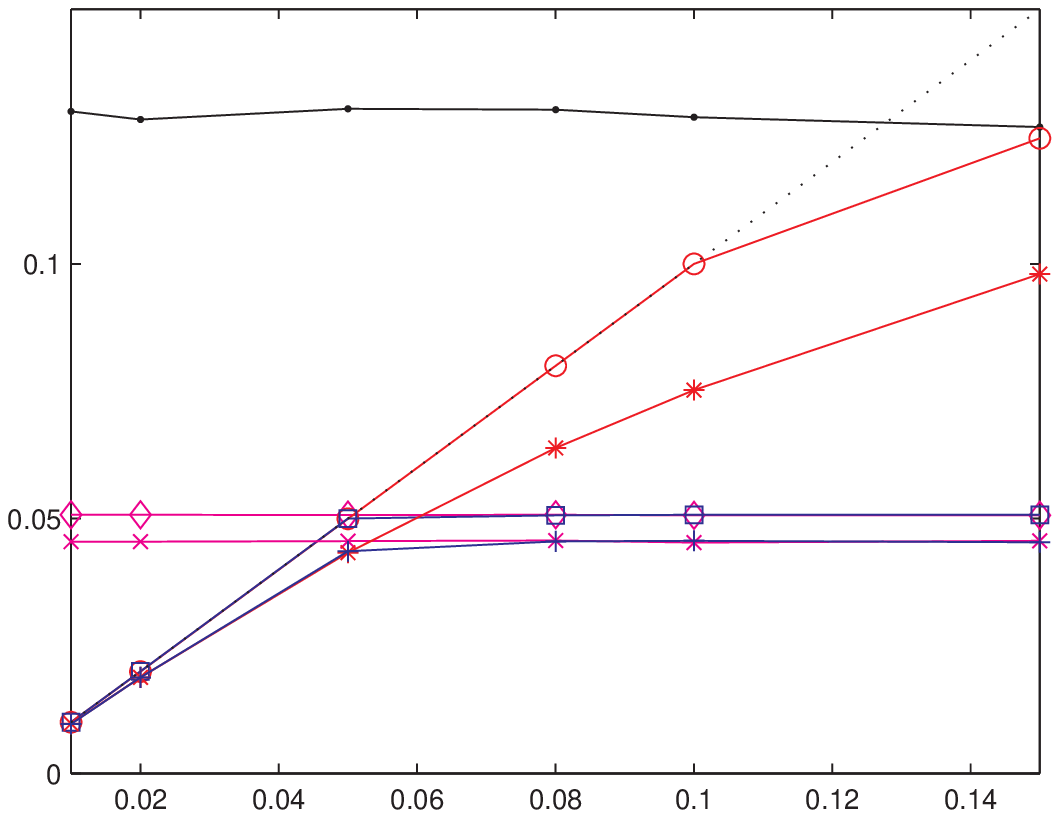}
\label{FigA:subfig4}}
\caption{Variation of $\bar{\varepsilon}_1$ w.r.t. $\bar{h}_{k,1}^m$, $\gamma_k$, $\check{p}_1$, and $\check{\varepsilon}_1$.}
\label{FigC:globfig}
\vspace{-.3cm}
\end{figure}

The main conclusions are: C1) Our schemes are always able to satisfy the constraints considered in each of the schemes. C2) The DSA long-term constraints achieve a better objective (sum capacity) than their short-term counterparts. Next we briefly elaborate on them. We begin by analyzing the feasibility claim. Figs. 2 and 3 confirm that the schemes always satisfy the constraints (small variations around the nominal value are due to the fact that the values plotted have been computed averaging over a \emph{finite} number of instants). Indeed, we observe that: ``None'' always violates the constraints; ``APC'' always satisfies both of them; ``AC'' always satisfies the long-term capacity loss constraint -Fig. 2-  and ``AP'' always satisfies the long-term interference power constraint -Fig. 3-; the schemes ``IPC'', ``IP'' and ``IC'' always oversatisfy the long-term constraints in Figs. 2 and 3. We also observe that ``AP'' and ``AC'' always satisfy the active constraint with equality (corroborating that they try to interfere the PUs as much as they are allowed to, so that the sum-rate of the SUs is as high as possible). We also observe that when the constraints are set to high (loose) values (see Figs. 2.c and 3.d), the performance of ``AP'' and ``AC'' (the schemes adhering to long-term constraints) coincides with that of the S1 (the scheme that ignores the DSA constraints). This indeed corroborates that our schemes are optimal. Moving to conclusion C2, the plots reveal that not only scheme ``APC'' performs always better than ``IPC'', but also that ``AP'' and ``AC'' perform better than ``IP'' and ``IC'', respectively. In other words, the schemes adhering to long-term DSA constrains always achieve a higher objective than their short-term counterparts. Intuitively, the long-term constraints allow SUs to interfere the PUs provided that the reward for the secondary network is high enough. This is referred to as ``cognitive diversity'' in \cite{CognitiveDiversity09,SurveyConvexCR_underlay_SPMag10}. The plots also reveal that the performance gap between the short-term and long-term formulations is larger when the scenario is more demanding.

\noindent\textbf{Test Case 2: imperfect CSI.} In this test case, we simulate incorporate imperfections to the CSI. The objective is threefold: O1) to numerically assess the performance (sum-capacity) loss due to the presence of CSI imperfections, O2) to show that our schemes are robust to CSI imperfections and adhere to the DSA constraints considered, and O3) to show that schemes that do not explicitly account for such imperfections either violate the DSA constraints or incur a significant loss of performance. Three different experiments are run. Only the APC and IPC schemes are simulated in this test case. The specific setup and the model for the CSI imperfections in each of the setups are described next.

In the first experiment, we consider that the CSI of the secondary network is quantized. The regions are designed using a scalar quantizer that splits the SNR domain into equally-probable regions. The results in Table \ref{T:Quantized} correspond to different quantization levels and demonstrate that for the average (APC) scheme, quantization of the CSI leads to small optimality loss w.r.t. the case of perfect CSI. Moreover, the resulting gap shrinks as the number of regions increases, being negligible when the number of regions is more than four (two feedback bits). The loss of optimality is more severe for the instantaneous (IPC) scheme. The reason is that none of the modes is activated during most of instants the PU is active.

\begin{table}
\caption{ {\small Variation of the number of quantization regions: $\check{\varepsilon}_{k,1}=5.0 \%$ and $\check{p}_{k,1}=0.20$.}} \label{T:Quantized}
\vspace{-0.25cm}
\begin{center}
\begin{tabular}{|c|c|c|c|c|c|c|}
\hline
\hline
& \multicolumn{3}{|c|}{APC} & \multicolumn{3}{|c|}{IPC} \\
\hline
$L$ & $\bar{c}_2$  & $\bar{\varepsilon}_1$ & $\bar{p}_1$ & $\bar{c}_2$  & $\bar{\varepsilon}_1$ & $\bar{p}_1$ \\
\hline
\hline
1 & 7.97 & 4.8 & 0.14  & 7.25 &  2.2 & 0.06 \\
\hline
2 & 12.41 & 5.0& 0.15  & 8.76 & 2.1 & 0.06\\
\hline
4 & 13.82 & 5.0 & 0.16 & 10.40 & 2.7 & 0.07 \\
\hline
8 & 14.66 & 5.0 & 0.15  & 10.48 & 2.5 & 0.07\\
\hline
$\infty$ &  15.16 & 5.0 & 0.16 &  14.45  & 4.0 & 0.12 \\
\hline
\hline
\end{tabular}
\vspace{-0.18205cm}
\end{center}
\vspace{-0.25cm}
\end{table}

In the second experiment, we assume that the information about the activity of the PUs is noisy and outdated. The time evolution of each $a_k[n]$ follows a Gilber-Elliot model with transition probabilities $P_{11}=0.975$, $P_{10}=0.025$, $P_{00}=0.9$, and $P_{01}=0.1$. Two sensing configurations are simulated. In the fist one, $P_{FA}=0.03$, $P_{MD}=0.02$, and the activity is measured every $N_a=5$ slots. In the second one, we set $P_{FA}=0.1$, $P_{MD}=0.1$ and $N_a=10$. We compare the performance of our schemes (3rd and 7th rows) with that of schemes: i) knowing the actual CSI, ii) ignoring the CSI imperfections, and iii) relying only on statistical CSI (labels ``-i'', ``-ii'' and ``-iii'' are used in the table). Clearly, as the sensor accuracy gets worse, the sum-capacity of the SUs gets smaller. The reason is simple, if the quality of the sensor is high, SUs can take advantage of time instants when the PUs are not present (in those instants the transmit power of the SU can be as high as they desire). Differently, when the quality of the sensors is poor, the SUs have to act as if the PUs were always present. This in turn implies that the loss due to sensing imperfections will be higher in those scenarios where the probability of the PUs being active is smaller (recall that we have set the probability of a PU being active to 80$\%$). Last but not least, we observe that our schemes always remain feasible even if the CSI contains imperfections. That is not the case if the schemes are implemented as if the CSI were perfect (see APC-ii and IPC-ii). Clearly, the sum-rate for APC-ii is higher than that of our scheme. The reason is that guaranteing the interference constraints with higher level of CSI requires more conservative transmission strategies.

\begin{table}
\caption{{\small Imperfections in the detection schemes: $\check{\varepsilon}_{k,1}=5.0 \%$ and $\check{p}_{k,1}=0.20$. Rows 3-6 correspond to $[P_{FA},P_{MD},N_a]=[0.02,$ $0.03,5]$.  Rows 7-10 correspond to $[P_{FA},P_{MD},N_a]=[0.1,0.1,10]$. }} \label{T:errors_sensing_activity}
\vspace{-0.15cm}
\begin{center}
\begin{tabular}{|c|c|c|c|c|c|c|}
\hline
\hline
& \multicolumn{3}{|c|}{APC} & \multicolumn{3}{|c|}{IPC} \\
\hline
\textsc{Version} & $\bar{c}_2$  & $\bar{\varepsilon}_1$ & $\bar{p}_1$ & $\bar{c}_2$  & $\bar{\varepsilon}_1$ & $\bar{p}_1$ \\
\hline
\hline
Optimal & 14.82 & 5.0 & 0.15 & 14.24 & 3.9 & 0.12\\
\hline
-i      & 15.18 & 5.0 & 0.15 & 14.46 & 4.3 & 0.13\\
\hline
-ii     & 15.22 & 5.5 & 0.17 & 14.51 & 8.7 & 0.17\\
\hline
-iii    & 14.39 & 4.3 & 0.15 & 13.57 & 3.1 & 0.09\\
\hline
\hline
Optimal & 14.54 & 5.0 & 0.15 & 13.80 & 3.3 & 0.10\\
\hline
-i      & 15.17 & 5.0 & 0.15 & 14.46 & 4.3 & 0.13\\
\hline
-ii     & 15.30 & 5.6 & 0.17 & 14.68 & 12.7 & 0.21\\
\hline
-iii    & 14.39 & 5.0 & 0.15 & 13.57 & 3.1 & 0.09\\
\hline
\hline
\end{tabular}
\vspace{-0.18205cm}
\end{center}
\vspace{-0.25cm}
\end{table}

Finally, the CSI of the SU-to-PU links is assumed to be noisy so that the ratio between the power of the true channel and the measurement noise is 4dB. As in the previous experiment, we simulate APC and IPC schemes and compare them with -i, -ii and -iii. Our schemes are feasible, and the achieved sum-rate is between the one obtained by the scheme that knows the actual CSI (-i) and the one that relies only on statistical CSI (-iii). Regarding the schemes ignoring the CSI imperfections, APC-ii achieves a slightly higher sum-rate that the scheme accounting for imperfections, but violates the interference constraints. We also observe that APC-ii achieves smaller sum-rate than APC-i, the reason being that the variance of the noisy channel is larger. The advantages are clearer in the instantaneous case: IPC-ii not only violates the constraints (this is not apparent in the table, which only lists average values), but also yields the worst performance [cf. the formulation in \eqref{E:optRA_c_inst_rate_loss_ind_users_imp}].

\begin{table}
\caption{{\small Imperfections in the CSI of the SU-to-PU links: $\check{\varepsilon}_{k,1}=5.0 \%$ and $\check{p}_1=0.15$.}}\label{T:errors_SNR_SU-to-PU_links}
\vspace{-0.15cm}
\begin{center}
\begin{tabular}{|c|c|c|c|c|c|c|}
\hline
\hline
& \multicolumn{3}{|c|}{APC} & \multicolumn{3}{|c|}{IPC} \\
\hline
\textsc{Version} & $\bar{c}_2$  & $\bar{\varepsilon}_1$ & $\bar{p}_1$ & $\bar{c}_2$  & $\bar{\varepsilon}_1$ & $\bar{p}_1$ \\
\hline
\hline
Optimal & 14.45 & 5.0 &0.15 & 8.68 & 3.0 & 0.08  \\
\hline
-i & 15.17 & 5.0 &0.15  & 14.46 & 4.2 & 0.12 \\
\hline
-ii & 14.50 & 5.8 &0.19  & 7.5 & 3.0 & 0.08 \\
\hline
-iii &  12.50 & 4.3 & 0.15 & 7.89 & 2.9 & 0.08 \\
\hline
\hline
\end{tabular}
\vspace{-0.18205cm}
\end{center}
\vspace{-0.15cm}
\end{table}

\vspace{-0.25cm}
\section{Conclusions}
\label{sec:concl}
\vspace{-0.057205cm}
This paper investigated the design of stochastic algorithms for CR scenarios with multiple primary and secondary users operating over time-varying (fading) channels. One of the most critical issues in CRs is how SUs coexist with (limit the interference to) PUs. Among the different metrics considered in the paper, the most important is the guarantee on the long-term (ergodic) capacity loss experienced by the PUs. Guaranteeing a certain rate for PUs is typically challenging because the presence of interference powers render the optimization non-convex. For the operating conditions considered in the paper we showed that two important facts hold. The first one is that the optimization problem which gives rise to the resource allocation schemes has zero duality gap, so that Lagrangian relaxation can be used without losing optimality. The second one is that in the dual domain the non-convex problem can be decoupled (separated) across channels and users. The latter implies that the optimization needs to be carried out only over a scalar variable, and thus enables implementation of efficient line-search algorithms. It was shown that the optimal resource allocation amounts to the maximization of a quality link functional which weights: the quality of the secondary links and the damage to the primary users. The terms in the quality link functional depend on the instantaneous CSI (which contains imperfections), and on several Lagrange multipliers (whose value depended on the long-term behavior of the system and the requirements of the primary and secondary networks). Simple stochastic algorithms that account for the imperfections in the sensing process are used to estimate and predict the actual value of the channel. Similarly, stochastic algorithms to estimate the optimum value of the multipliers online were also developed. Future work includes consideration of multiple antenna, development of distributed (including multi-hop) implementations, and joint design of the sensing and resource allocation schemes.

\section*{Appendix A: On the optimality of the RA}

As pointed out in Sec. \ref{S:ProblemFormulation}, there are three sources of nonconvexity in \eqref{E:optRA}: \textbf{i)} scheduling coefficients $w_{k,2}^m$ are constrained to belong to the non-convex set $\{0,1\}$; \textbf{ii)} monomials $w_{k,2}^m p_{k,2}^m$, $w_{k,2}^m r_{k,2}^m$, and $w_{k,2}^m r_{k,1}$ are not jointly convex; and \textbf{iii)} constraints \eqref{E:optRA_c_av_rate_loss} are not convex w.r.t. $p_{k,2}^m$. In this appendix, we first discuss how the two first sources of non-convexity can be bypassed. Then, we analyze why the reformulated problem has zero-duality gap. Finally, we show that the RA in \eqref{E:opt_ind_case1}-\eqref{E:opt_sched_case1} is optimum.

The way do deal with \textbf{i)} is to relax $w_{k,2}^m\in\{0,1\}$ and consider $w_{k,2}^m\in[0,1]$. In general, such a relaxation will give rise to solutions $w_{k,2}^{m*}$ that do not satisfy the original constraint $w_{k,2}^m\in\{0,1\}$. However, it can be shown that if $w_{k,2}^m\in\{0,1\}$ is replaced with $w_{k,2}^m\in[0,1]$, the solution of \eqref{E:optRA} satisfies
$w_{k,2}^{m*}\in\{0,1\}$ with probability one. This easily follows from the expression for $w_{k,2}^{m*}$ in \eqref{E:opt_sched_case1}, which was derived considering $w_{k,2}^m\in[0,1]$. Clearly, \eqref{E:opt_sched_case1} dictates that $w_{k,2}^{m*}$ is either zero or one. The only problem arises if there are two SUs $m_1$ and $m_2$ with positive transmit power satisfying $\varphi_k^{m_1}(p_k^{m_1*}[n]) = \varphi_k^{m_2}(p_k^{m_2*}[n]) = \max_l \varphi_k^l(p_k^{l*}[n])$. Since $\varphi_k^l$ and $p_k^{l*}$ are continuous functions of several (continuous) random variables, the probability of that event is zero. For further details on this specific issue, we refer the reader to the end of this appendix, where the optimal scheduling is found [cf. \eqref{E:App_opt_sched_problem}]. Nonetheless, it is worth clarifying that from a practical perspective, the problems associated with the event of two users achieving the same indicator (which happens if, for example, the channel is a discrete random process) can be easily bypassed. For example, by using smooth scheduling approximations, which are asymptotically optimal; see \cite{amggjr_tsp11} for details.

To deal with \textbf{ii)} we follow the same approach used in other RA problems; see, e.g., \cite{amggjr_tsp11}. The idea is to define auxiliary (dummy) variables $\tilde{p}_{k,2}^m:=w_{k,2}^m p_{k,2}^m$. The problem in \eqref{E:optRA} is then reformulated replacing $p_{k,2}^m$ with $\tilde{p}_{k,2}^m/w_{k,2}^m$. After straightforward mathematical manipulations, it can be shown that: a) the non-convexity caused by the monomials is indeed solved and b) the reformulated problem yields the same (Karush-Kuhn-Tucker) KKT conditions than those of the original \eqref{E:optRA}. More specifically, the only difference between the solution of \eqref{E:optRA} considering the original variables and the one considering the dummy variables are the values of $p_{k,2}^{m*}$ for users $m$ such that $w_{k,2}^{m*}=0$. Clearly, such a difference is irrelevant from a performance perspective and hence, the optimization can be carried out using any of them.

Regarding the zero duality gap in \textbf{iii)}, the basic idea is that the source of non-convexity comes from a constraint of the form $\mathbbm{E}_{\mathbf{x}}[g(\mathbf{y},\mathbf{x})]$, where $g(\mathbf{y},\mathbf{x})$ is a non-convex function w.r.t. $\mathbf{y}$, and $\mathbf{x}$ is a continuous random process with infinite support. Here $\mathbf{y}$ is the power; $\mathbf{x}$ is the CSI; and $g(\mathbf{y},\mathbf{x})$ is the expression for the instantaneous capacity, i.e.  $\log_2(1+\gamma_{k,1}/(1+h_{k,1}^m[n]p_{k,2}^m[n])$. The proof is omitted due to space limitations, but we refer the reader to either \cite{AleGG10ZeroDualityGap}, or \cite[App. A]{KetanNikosZeroDualGap} for further details.

To derive the optimum RA in \eqref{E:opt_ind_case1}-\eqref{E:opt_sched_case1} we start by writing the Lagrangian of \eqref{E:optRA}. To do so, let $\mathbf{z}$ be a vector containing all primal variables: $w_{k,2}^m(\mathbf{h})$, $p_{k,2}^m(\mathbf{h})$ $\forall (k, m, \mathbf{h})$. Note that $\mathbf{z}$ has infinite length because $\mathbf{h}$ takes infinite values. Moreover, let $\boldsymbol{\lambda}$ be a vector containing all dual  variables (multipliers): $\pi^m$, $\theta_k$, $\rho_k$ $\forall (k,m)$. The Lagrangian is then
\begin{multline}\label{E:App_Def_Lagr}
\mathcal{L} (
\mathbf{z}, \boldsymbol{\lambda}
) = \mathbb{E}_{\mathbf{h}}\Big[\biggl(\sum_{m,k} \beta^m w_{k,2}^m(\mathbf{h})r_{k,2}^m(h_{k,2}^mp_{k,2}^m(\mathbf{h}))\biggr)\\
-\sum_m\pi^m \biggl(\sum_k w_{k,2}^m(\mathbf{h}) p_{k,2}^m(\mathbf{h}) - \check{p}_{m,2}\biggr)\\
-\sum_k\theta_k a_{k,1}\biggl(\sum_m w_{k,2}^m(\mathbf{h}) h_{k,1}^m p_{k,2}^m(\mathbf{h})- \check{p}_{k,1}	 \biggr) \\
 +  \sum_k\rho_k a_{k,1}\biggl(\sum_m w_{k,2}^m(\mathbf{h}) r_{k,1}(h_{k,1}^m p_{k,2}^m(\mathbf{h})) - \check{r}_{k,1}
	\biggr)\Big].
\end{multline}
For a given $\boldsymbol{\lambda}$, we need to maximize $\mathcal{L}(\mathbf{z}, \boldsymbol{\lambda}
)$ w.r.t. $\mathbf{z}$ and guarantee that the solution satisfies the short-term constraints in \eqref{E:optRA_inst_const}. The structure of $\mathcal{L}(\mathbf{z}, \boldsymbol{\lambda})$ and the constraints allows for a separate optimization w.r.t. $w_{k,2}^m(\mathbf{h})$ and $p_{k,2}^m(\mathbf{h})$. First we will find an expression for $p_{k,2}^{m*}(\mathbf{h},\boldsymbol{\lambda})$ which holds for any value of $w_{k,2}^m(\mathbf{h})$. Then, we will use $p_{k,2}^{m*}(\mathbf{h},\boldsymbol{\lambda})$ to find $w_{k,2}^{m*}(\mathbf{h},\boldsymbol{\lambda})$.

To handle perfect and imperfect CSI jointly, the expectation in \eqref{E:App_Def_Lagr} is written as $\mathbb{E}_{\mathbf{h}}[\cdot]=\mathbb{E}_{\mathbf{\tilde{h}}}[\mathbb{E}_{\mathbf{b}(x|\mathbf{\tilde{h}})}[\cdot]]$, so that
\begin{multline}\label{E:App_Def_Lagr2}
\hspace{-.4cm}\mathcal{L} (\mathbf{z}, \boldsymbol{\lambda}) =
\sum_m \pi^m\check{p}_{m,2} + \sum_k \mathbb{E}_{\mathbf{\tilde{h}}}[\mathbb{E}_{\mathbf{b}(x|\mathbf{\tilde{h}})}[a_{k,1}(\theta_k\check{p}_{k,1} -\rho_k\check{r}_{k,1})]]
\\
+\mathbb{E}_{\mathbf{\tilde{h}}}\biggl[\sum_{m,k} w_{k,2}^m(\mathbf{\tilde{h}})
\mathbb{E}_{\mathbf{b}(x|\mathbf{\tilde{h}})}\Big[\beta^m r_{k,2}^m(h_{k,2}^mp_{k,2}^m(\mathbf{\tilde{h}}))-\pi^m p_{k,2}^m(\mathbf{\tilde{h}})\\
\hspace{-.2cm}-\theta_k a_{k,1} h_{k,1}^m p_{k,2}^m(\mathbf{\tilde{h}})
 +  \rho_k a_{k,1} r_{k,1}(h_{k,1}^m p_{k,2}^m(\mathbf{\tilde{h}})) \Big]\biggr].\hspace{.6cm}
\end{multline}
Clearly, when the CSI is perfect, the inner expectation is not needed and can be dropped. Taking into account that the two first terms in \eqref{E:App_Def_Lagr2} do not depend on $\mathbf{z}$, and using the definition of the link quality indicator $\tilde{\varphi}_k^m$ in \eqref{E:opt_ind_case1_imp}, maximizing $\mathcal{L} (\mathbf{z}, \boldsymbol{\lambda})$ w.r.t. $\mathbf{z}$ amounts to maximize
\begin{equation}\label{E:App_Def_Lagr3}
\mathcal{L}' (\mathbf{z}, \boldsymbol{\lambda}) :=
\mathbb{E}_{\mathbf{\tilde{h}}}\left[\sum_{m,k} w_{k,2}^m(\mathbf{\tilde{h}})\tilde{\varphi}_k^m(\mathbf{\tilde{h}},\boldsymbol{\lambda},p_{k,2}^m(\mathbf{\tilde{h}}))\right].
\end{equation}
w.r.t. $\mathbf{z}$. Clearly, the \emph{unconstrained} maximization of $\mathcal{L}'(\mathbf{z}, \boldsymbol{\lambda})$ can be performed separately for each of the $(m,k,\mathbf{\tilde{h}})$ terms. However, the optimal solution also needs to satisfy the instantaneous constraints in \eqref{E:optRA_inst_const}, namely: $\sum_m w_{k,2}^m(\mathbf{\tilde{h}})= 1$; $0\leq w_{k,2}^m(\mathbf{\tilde{h}}) \leq 1$;  and  $0\leq p_{k,2}^m(\mathbf{\tilde{h}}) \leq \check{p}_{k,2}^m(\mathbf{\tilde{h}})$. Indeed, since the instantaneous constraints on $p_{k,2}^m(\mathbf{\tilde{h}})$ are decoupled across $m$, $k$ and $\mathbf{\tilde{h}}$, the optimization \emph{over the power} can be performed separately for each of the $(m,k,\mathbf{\tilde{h}})$ terms. To find $p_{k,2}^{m*}(\mathbf{\tilde{h}},\boldsymbol{\lambda})$ we consider two different cases: i) if $w_{k,2}^m(\mathbf{\tilde{h}})>0$, then the optimum $p_{k,2}^{m*}(\mathbf{\tilde{h}},\boldsymbol{\lambda})$ is found by maximizing $\tilde{\varphi}_k^m(\mathbf{\tilde{h}},\boldsymbol{\lambda},p_{k,2}^m(\mathbf{\tilde{h}}))$ and projecting the solution onto the feasible interval $[0,\check{p}_{k,2}^m(\mathbf{\tilde{h}})]$; and ii) if $w_{k,2}^m(\mathbf{\tilde{h}})=0$, then any value of $p_{k,2}^m(\mathbf{\tilde{h}})$ is equally optimum, including the one which is optimum for i). As a result, we can conclude that finding $p_{k,2}^{m*}(\mathbf{\tilde{h}},\boldsymbol{\lambda})$ by maximizing $\tilde{\varphi}_k^m(\mathbf{\tilde{h}},p_{k,2}^m(\mathbf{\tilde{h}},\boldsymbol{\lambda}))$ and projecting onto $[0,\check{p}_{k,2}^m(\mathbf{\tilde{h}})]$ is optimum for any value of $w_{k,2}^m(\mathbf{\tilde{h}})$. This is indeed the result in \eqref{E:opt_pow_case1} and \eqref{E:opt_pow_case1_imp} for the cases of perfect and imperfect CSI, respectively.

Once $p_{k,2}^{m*}(\mathbf{\tilde{h}},\boldsymbol{\lambda})$ are known $\forall (k,m)$, we are ready to find $w_{k,2}^{m*}(\mathbf{\tilde{h}},\boldsymbol{\lambda})$. To carry out this task, we substitute $p_{k,2}^{m*}(\mathbf{\tilde{h}},\boldsymbol{\lambda})$ into \eqref{E:App_Def_Lagr2} and rely on the fact that the short-term scheduling constraints are decoupled across channels. As a result, for each $\mathbf{\tilde{h}}$, it suffices to solve $K$ instances (one per $k$) of
\begin{subequations}
\label{E:App_opt_sched_problem}
\begin{align}
\underset{\{w_{k,2}^m(\mathbf{\tilde{h}})\}_{m=0}^M}{\max}& ~~\sum_{m=0}^M w_{k,2}^m(\mathbf{\tilde{h}}) \tilde{\varphi}_k^m(\mathbf{\tilde{h}},\boldsymbol{\lambda},p_{k,2}^{m*}(\mathbf{\tilde{h}})))&  \\
\mathrm{s.~to}: \quad & ~~\sum_{m=0}^M  w_{k,2}^m(\mathbf{\tilde{h}}) = 1 &  \label{E:App_wleq1x}\\
 &~~0\leq w_m^k (\mathbf{\tilde{h}})\leq 1 ~~~~\forall m & \label{E:App_wgeq0x}
\end{align}
\end{subequations}
whose solution yields $\{w_{k,2}^{m*}(\mathbf{\tilde{h}},\boldsymbol{\lambda})\}_{m=0}^M$. Since \eqref{E:App_opt_sched_problem} is linear in $w_{k,2}^m(\mathbf{\tilde{h}})$, the solution is straightforward and consists of setting $w_{k,2}^{m*}(\mathbf{\tilde{h}},\boldsymbol{\lambda})=1$ for the user $m$ which maximizes $\tilde{\varphi}_k^m(\mathbf{\tilde{h}},\boldsymbol{\lambda},p_{k,2}^{m*}(\mathbf{\tilde{h}},\boldsymbol{\lambda}))$, while setting $w_{k,2}^{m*}(\mathbf{\tilde{h}},\boldsymbol{\lambda})=0$ for all other users. If the winner user is unique, this policy can be written in closed form using the indicator function as $w_{k,2}^{m*}(\mathbf{\tilde{h}},\boldsymbol{\lambda})=\mathbbm{1}_{\{ (m=\arg\max_{l} \tilde{\varphi}_k^l(p_k^{l*}(\mathbf{\tilde{h}},\boldsymbol{\lambda})))\}}$. If more than one user attains the maximum (this event will be referred to as a tie), choosing any of them is optimum from the point of view of \eqref{E:App_opt_sched_problem}. However, since $\tilde{\varphi}_k^m(\mathbf{\tilde{h}},\boldsymbol{\lambda},p_{k,2}^{m*}(\mathbf{\tilde{h}},\boldsymbol{\lambda}))$ is a continuous non-negative random variable, ties in practice only occur if $p_{k,2}^{m*}(\mathbf{\tilde{h}},\boldsymbol{\lambda})=0$ for all $m$. In such a case, the LQI is the same for all $M+1$ users and any of them could be selected. In this situation, we assign the access to the virtual user $m=0$, i.e. we set $w_{k,2}^{m*}(\mathbf{\tilde{h}},\boldsymbol{\lambda})=0$ for all $m>0$. Combining these two conditions we can write $w_{k,2}^{m*}(\mathbf{\tilde{h}},\boldsymbol{\lambda})=\mathbbm{1}_{\{ (m=\arg\max_{l} \tilde{\varphi}_k^l(p_k^{l*}(\mathbf{\tilde{h}},\boldsymbol{\lambda})))\}}\mathbbm{1}_{\{(p_{k,2}^{m*}(\mathbf{\tilde{h}},\boldsymbol{\lambda})>0~\vee~ m=0)\}}$ for all $m>0$. This is precisely the solution in \eqref{E:opt_sched_case1} and \eqref{E:opt_sched_case1_imp}, for the cases of perfect and imperfect CSI, respectively


\bibliographystyle{IEEEbib}

\begin{thebibliography}{1}


\bibitem{Imperf_CRref4}
K. S. Ahn and R. W. Heath, ``Performance analysis of maximum ratio combining with imperfect channel estimation in the presence of cochannel interferences,'' \emph{IEEE Trans. Wireless Commun.}, vol. 8, no. 3, pp. 1080–-1085, Mar. 2009.


\bibitem{Alouini11}
J. A. Ayala Solares, Z. Rezki, and M.S. Alouini, ``Optimal power allocation of a sensor node under different rate constraints,'' \emph{Proc. of IEEE Intl. Conf. on Commun.}, Otawa, Canada, Jun. 10--15, 2012.



\bibitem{Bertsekas_Book_CA03} D. Bertsekas, A. Nedic, and A. E. Ozdaglar, {\em Convex Analysis and Optimization}, Athena Scientific, 2003.


\bibitem{QZhao08}
Y. Chen, Q. Zhao, and A. Swami, ``Joint design and separation principle for opportunistic spectrum access in the presence of sensing errors,'' \emph{IEEE Trans. Inf. Theory}, vol. 54, no. 5, pp. 2053–-2071, May 2008.

\bibitem{emiliano11twc}
E. Dall'Anese, S.-J. Kim, G. B. Giannakis, and S. Pupolin, ``Power control for cognitive radio networks under channel uncertainty,'' \emph{IEEE Trans. Wireless Commun.}, vol. 10, no. 10, pp.
3541--3551, Oct. 2011.


\bibitem{DiLorenzoBarbarossa_tsp11}
P. {Di Lorenzo} and S. Barbarossa, ``A bio-inspired swarming algorithm for decentralized access in cognitive radio,'' \emph{IEEE Trans. Signal Process.}, vol. 59, no. 12, pp. 6160--6174, Dec. 2011.

\bibitem{GatsisErrorsExchanges}
N. Gatsis and G. B. Giannakis, ``Power control with imperfect exchanges and applications to spectrum sharing,'' \emph{IEEE Trans. Signal Process.}, vol. 59, no. 7, pp. 3410--3423, Jul. 2011.


\bibitem{GatsisDSA}
N. Gatsis, A. G. Marques, and G. B. Giannakis, ``Power control for cooperative dynamic spectrum access networks with diverse QoS constraints,'' \emph{IEEE Trans. Commun.}, vol. 58, no. 3, pp. 933--944, Mar. 2010.


\bibitem{GoldsmithBook}
A. Goldsmith, {\em Wireless Communications}, Cambridge Univ. Press, 2005.

\bibitem{GoldsmithCR09}
A. Goldsmith, S. A. Jafar, I. Maric and S. Srinivasa, ``Breaking spectrum gridlock with cognitive radios: An information theoretic perspective,'' \emph{Proc. IEEE}, vol. 97, no. 5, pp. 894--914, May 2009.

\bibitem{Sergiy_icassp11}
X. Gong, S. Vorobyov, and C. Tellambura, ``Optimal bandwidth and power allocation for sum ergodic capacity under fading channels in cognitive radio networks'', \emph{IEEE Trans. Signal Process.}, vol. 59, no. 4, pp 1814--1826, Apr. 2011.

\bibitem{Haykin05}
S. Haykin, ``Cognitive radio: Brain-empowered wireless communications,'' {\em IEEE J. Sel. Areas Commun.}, vol. 23, no. 2, pp. 201--220, Feb. 2005.




\bibitem{RA_CR_Kang09}
X. Kang, Y. Liang, A. Nallanathan, H.K. Garg and R. Zhang, ``Optimal power allocation for fading channels in cognitive radio networks: ergodic capacity and outage capacity,'' \emph{IEEE Trans. Wireless Commun.}, vol. 8, no. 2, pp. 940--950, Feb. 2009.


\bibitem{SINR1}
D.-I. Kim, L.-B. Le, and E. Hossain, ``Joint rate and power allocation for cognitive radios in dynamic spectrum access environment,'' \emph{IEEE Trans. Wireless Commun.}, vol. 7, no. 12, pp.
5517--5527, Dec. 2011.


\bibitem{SI_jsac_lrf}
D. J. Love, R. W. Heath, V. K. Lau, D. Gesbert, B. Rao, and M. Andrews, ``An overview of limited feedback in wireless communication systems,'' \emph{IEEE J. Sel. Areas Commun.}, vol. 26,  no. 8, pp. 1341--1365,  Aug. 2008.


\bibitem{amggjr_tsp11}
A. G. Marques, G. B. Giannakis, and J. Ramos, ``Optimizing orthogonal multiple access based on quantized channel state information,'' \emph{IEEE Trans. Signal Process.}, vol. 59, no. 10, pp.  5023--5038, Oct. 2011.

\bibitem{JSAC}
A. G. Marques, L. M. Lopez-Ramos, G. B. Giannakis, and J. Ramos, ``Resource allocation for interweave and underlay {CRs} under probability-of-interference constraints'', \emph{IEEE J. Sel. Areas Commun.}, vol. 60, no. 10, pp. (to appear), Nov. 2012.

\bibitem{am_etal_tvt12}
A. G. Marques, L. M. Lopez-Ramos, G. B. Giannakis, J. Ramos, and A. Caamano, ``Optimal cross-layer resource allocation in cellular networks using channel and queue state information,'' \emph{IEEE Trans. Vehic. Tech.}, vol. 61, no. 6, pp. 2789--2807, Jul. 2012.

\bibitem{quantizedCR}
A. G. Marques, X. Wang, and G. B. Giannakis, ``Dynamic resource management for cognitive radios using limited-rate feedback,'' \emph{IEEE Trans. Signal Process.}, vol. 57, no. 9, pp. 3651--3666, Sep. 2009.

\bibitem{refInterf2}
L. Musavian and S. Aissa, ``Fundamental capacity limits of cognitive radio in fading environments with imperfect channel information,'' \emph{IEEE Trans. Commun.}, vol. 57, no. 11, pp. 3472--3480, Nov. 2009.


\bibitem{CR_jsac_GameTheory}
D. Niyato and E. Hossain, ``Competitive pricing for spectrum sharing in cognitive radio networks: dynamic game, inefficiency of Nash equilibrium, and collusion,'' \emph{IEEE J. Sel. Areas Commun.}, vol. 26,  no. 1, pp. 192--202,  Jan. 2008.

\bibitem{KetanNikosZeroDualGap}
K. Rajawat, N. Gatsis, and G. B. Giannakis, ``Cross-layer designs in coded wireless fading networks with multicast,'' \emph{IEEE/AMC Trans. Networking}, vol. 19, no. 5, pp. 1276--1289, Oct. 2011.

\bibitem{AleTSP11}
A. Ribeiro, ``Ergodic stochastic optimization algorithms for wireless communication and networking,'' \emph{IEEE Trans.  Signal Process.}, vol. 58, no. 12, pp. 6369--6386, Dec. 2010.


\bibitem{AleGG10ZeroDualityGap}
A. Ribeiro and G. B. Giannakis, ``Separation principles in wireless networking,'' \emph{IEEE Trans. Inf. Theory}, vol. 56, no. 9, pp. 4488--4505, Sep. 2010.

\bibitem{Imperf_CRref2}
H. A. Suraweera, P. J. Smith, and M. Shafi, ``Capacity limits and performance analysis of cognitive radio with imperfect channel knowledge,'' \emph{IEEE Trans. Vehic. Tech.}, vol. 59, no. 4, pp. 1811--1822 , May 2010.


\bibitem{NeelyCR09}
R. Urgaonkar and M. Neely, ``Opportunistic scheduling with reliability guarantees in cognitive radio networks,'' \emph{IEEE Trans. Mobile Comp.}, vol. 8, no. 6, pp. 766--777, Jun. 2009.


\bibitem{Joint_S_RA_Xin10}
X. Wang, ``Joint sensing-channel selection and power control for cognitive radios,'' \emph{IEEE Trans. Wireless Commun.}, vol. 10, no. 3, pp. 958--967, Mar. 2011.

\bibitem{XinInfoTheory}
X. Wang and G. B. Giannakis, ``Power-efficient resource allocation in time-division multiple access over fading channels,'' \emph{IEEE Trans. Inf. Theory}, vol. 54, no. 3, pp. 1225--1240, Mar. 2008.



\bibitem{SurveyConvexCR_underlay_SPMag10}
R. Zhang, Y.-C. Liang, and S. Cui, ``Dynamic resource allocation in cognite radio networks: a convex optimization perspective,'' \emph{IEEE Signal Process. Mag.}, vol. 27, no. 5, pp. 102--114, May 2010.

\bibitem{CognitiveDiversity09}
R. Zhang, ``On peak versus average interference power constraints for protecting primary users in cognitive radio networks,'' \emph{IEEE Trans. Wireless Commun.}, vol. 8, no. 4, pp. 2112-–2120, Apr. 2009.

\bibitem{SurvZhaoSadler07}
Q. Zhao and B. M. Sadler, ``A survey of dynamic spectrum access,'' \emph{IEEE Signal Process. Mag.}, vol. 24, no. 3, pp. 79–-89, May 2007.







\end{thebibliography}

\end{document}